\pdfoutput=1 

\documentclass[prx, twocolumn, nofootinbib, floatfix]{revtex4-2}

\usepackage{CJK}
\usepackage{amssymb}
\usepackage{amsmath}
\usepackage{mathtools}
\usepackage{amsfonts}
\usepackage{float}
\usepackage{textcmds}
\usepackage{gensymb}
\usepackage[dvipsnames]{xcolor}
\usepackage{textcomp}
\usepackage{nicefrac}
\usepackage{comment}
\usepackage{multirow}
\usepackage{setspace}
\usepackage{graphicx}\usepackage{dcolumn}\usepackage{bm}

\graphicspath{ {./fig/} }

\usepackage[justification=centerlast]{caption} \usepackage[labelformat=parens,subrefformat=parens,font=smaller]{subcaption} 

\usepackage{xargs}

\usepackage{etoolbox}

\usepackage[hidelinks]{hyperref}
\usepackage{cleveref}
\usepackage{orcidlink}

\usepackage[T1]{fontenc}
\usepackage[english]{babel}

\usepackage{mathrsfs,pgf,pgfplots,tikz,pgfplotstable}
\usetikzlibrary{shapes, positioning, chains, arrows, decorations.text, decorations.markings,
    arrows.meta, calc, shapes.geometric, angles, quotes, intersections,
    datavisualization.formats.functions, datavisualization.polar, external, fadings}

\usepgfplotslibrary{polar}

\pgfplotsset{compat=1.18}
\pgfplotsset{every axis/.append style={
            ticklabel style = {font=\small},
            ylabel near ticks,
            label style = {font=\small},
        }}

\tikzexternaldisable

\def\affETHZ{Institute of Geophysics, ETH Zürich, Switzerland}
 \colorlet{colToDo}{YellowOrange!35}
\colorlet{colInfo}{CornflowerBlue!10}
\colorlet{colquestion}{RedOrange!35}

\colorlet{colScat}{black!8}
\colorlet{colScatBoundary}{black!40}

\colorlet{colAnotations}{gray!70}

\presetkeys{todonotes}{linecolor=colToDo,backgroundcolor=colToDo,bordercolor=white}{}

\newcommandx{\info}[2][1=]{\todo[={info},inline,linecolor=colInfo,backgroundcolor=colInfo,bordercolor=colInfo,#1]{#2}}
\renewcommand\info[2][]{} \newcommandx{\question}[2][1=]{\todo[caption={question},linecolor=colquestion,backgroundcolor=colquestion,bordercolor=colquestion,#1]{#2}}

\crefname{figure}{Fig.}{Figs.}
\Crefname{eqs}{Equation}{Equations}

\tikzstyle{decision} = [diamond, draw, text width=4.75em, text badly centered, inner sep=0pt]
\tikzstyle{block} = [rectangle, draw, text width=21em, text centered, rounded corners, minimum height=1.75em]
\tikzstyle{line} = [draw, -latex']
\tikzstyle{cloud} = [draw, ellipse, node distance=3cm, minimum height=2em]
\tikzstyle{arrow} = [thick,->,>=stealth]

\tikzset{
recShape/.style = { isosceles triangle, isosceles triangle apex angle=60, rotate=270,fill=\colrec },
ray/.style = { -{Latex},shorten <=2pt,shorten >=2pt,black!90 },
bw/.style = {thick},
rec/.style = { \colrec,dotted, very thick },
src/.style = { \colsrc,dotted, very thick },
scat/.style = {diamond, draw = none, fill = gray!40, minimum width = \scatL, minimum height = \scatW},
sourceStar/.style = {star, fill=\colsrc, star point ratio=2.2, scale=0.43},
sizeBlock/.style={minimum height=\baselineskip*1.7,
		text width = 0.87*\columnwidth},
startstop/.style={rounded corners,
		text centered,
		draw=red!30,
		thick,
		fill=red!14},
data/.style={trapezium,
		trapezium left angle=85,
		trapezium right angle=95,
		trapezium stretches body=true,
		text centered,
		draw=blue!20,
		thick,
		fill=blue!9,
		sizeBlock},
decision/.style={
		diamond,
		aspect=2.5,
		text centered,
		draw=green!25,
		thick,
		fill=green!10,
		text width=4.5em,
		inner sep=1pt},
action/.style={rectangle,
		text centered,
		draw=orange!35,
		thick,
		fill=orange!14,
		text width=5em,
		align=center,
		sizeBlock},
tikzPlot/.style={trim axis left, trim axis right,inner frame sep=0}
}

\newcounter{cnt}
\newcommand{\cdef}[1]{\stepcounter{cnt}\xglobal \definecolor{#1}{HTML}{#1}
\expandafter\newcommand\csname c\Roman{cnt}\endcsname{#1}}

\cdef{1f77b4}
\cdef{ff7f0e}
\cdef{2ca02c}
\cdef{d62728}
\cdef{9467bd}
\cdef{8c564b}
\cdef{e377c2}
\cdef{7f7f7f}
\cdef{bcbd22}
\cdef{17becf}

\newcommand{\colrec}{\cI}
\newcommand{\colsrc}{\cIV} \newcommand{\colscat}{\cVIII!70!white}
\newcommand{\colscatibc}{\cVIII!90!white}

\newcommand{\so}{S^{\text{O}}}
\newcommand{\si}{S^{\text{I}}}
\newcommand{\xin}{\bm{x}^{\text{I}}}
\newcommand{\xout}{\bm{x}^{\text{O}}}
\newcommand{\gf}{Green's function}
\newcommand{\gfs}{Green's functions}

\newcommand{\gpq}{G^{p\mid q}}
\newcommand{\gpf}{G^{p\mid f}}

\newcommand{\pluseq}{\mathrel{{+}{=}}}

\newcommand{\monopole}{
	\tikz[baseline=-0.55ex]{
		\def\rMono{1ex/2.9};
		\draw[fill=black!90] (0,-1pt)  circle [radius=\rMono];
		\draw[fill=black!90] (0,2*\rMono)  circle [radius=\rMono];
	}
}

\newcommand{\dipole}{
	\tikz[baseline=-0.55ex]{
		\def\rMono{1ex/2.9};
		\draw[fill=white] (0,-1pt)  circle [radius=\rMono];
		\draw[fill=black!90] (0,2*\rMono)  circle [radius=\rMono];
	}
}

\addto\captionsenglish{}
\newlength{\figHResults}
\newlength{\figWResults}
\newlength{\rInsetResults}

\global \deflength{\figHResults}{7cm}
\global \deflength{\figWResults}{(\textwidth)/2/100*49} 

\global \deflength{\rInsetResults}{\figWResults/2-\figWResults/8}

\pgfdeclareradialshading{tikz@lib@fade@circle@50}{\pgfpointorigin}{color(0pt)=(pgftransparent!0); color(10bp)=(pgftransparent!0);color(25bp)=(pgftransparent!100); color(50bp)=(pgftransparent!100)}
\pgfdeclarefading{circle with fuzzy edge 50 percent}{\pgfuseshading{tikz@lib@fade@circle@50}} 	
\begin{document}

\preprint{EEG}

\title{Acoustic Cloning}\thanks{Author-prepared version of the article published as J.~M\"uller \textit{et~al.}, Phys.\ Rev.\ Applied \textbf{20}, 064014 (2023), \href{https://doi.org/10.1103/PhysRevApplied.20.064014}{doi:10.1103/PhysRevApplied.20.064014}. \copyright{} 2023 American Physical Society.}

\author{Jonas Müller\orcidlink{0000-0003-4530-2486}}
\email{jmuller.research@gmail.com}

\author{Theodor S. Becker\orcidlink{0000-0003-0065-359X}}
\author{Xun Li\orcidlink{0000-0002-2104-5182}}\author{Johannes Aichele\orcidlink{0000-0001-8019-9053}}
\author{Marc Serra-Garcia\orcidlink{0000-0002-4136-253X}}
\author{Johan O. A. Robertsson\orcidlink{0000-0002-3292-385X}}
\author{Dirk-Jan van Manen\orcidlink{0000-0002-8250-2826}}

\affiliation{\affETHZ}

\date[]{}

\begin{abstract}
Cloning refers to producing identical copies of existing objects. Here, we experimentally show how to clone acoustic scattering objects. We acquire a digital twin and bring it back to life - a simple two-step process. First, we use broadband speakers to illuminate the scattering object within a closed receiver aperture. From these recorded reverberative data, we retrieve the object's scattering \gfs{} using multidimensional  deconvolution. In the second step, the acoustic scatterer is holographically reconstructed using the acquired scattering \gfs{}. The hologram scatters any wavefield in real-time exactly like the original object would. Low-latency feedback reproduces all orders of interactions between the physical wavefield and the numerically defined hologram. This two-step process is demonstrated by cloning and modifying several rigid scatterers in a two-dimensional acoustic waveguide. Applications range from fully realistic digital scattering models to efficient metamaterial experimentation.
\end{abstract}

{\tikzexternaldisable
\maketitle
}

\section{Introduction}
\label{sec:intro}

Acoustic scattering constitutes the physical basis for a wide range of fields, including room acoustics, seismic, ultrasound, sonar imaging, and the design of acoustic metamaterials. 
In these examples, acoustic waves can carry information on the object that scatters the waves (imaging) or constitute a desired design feature of the object itself (metamaterial).

Acoustic cloning, which we introduce in this work, signifies the replication of an acoustic scatterer. That is, the reproduction of all interactions between an object and an arbitrary acoustic wavefield. Any wavefield probing the clone will still be subjected to the object's scattering, even if the physical scatterer is absent.

The reproduction of a scattering wavefield was first introduced in optics. Gabor invented the concept of holography~\cite{gaborHolography194819711972}, which led to important developments in optical imaging over the last decades, for instance, in microscopy \cite{de_la_torre_i_gabors_2022,hariharan_basics_2002}. The classical framework of holography aims at reproducing the scattered wavefield on a holographic plane, such that further imaging of the hologram over different angles remains possible. It generally focuses on imaging the scattered wavefield of an object over a limited aperture and at a single frequency. Holography has been adapted in near-field acoustics \cite{maynardNearfieldAcousticHolography1985}, where array measurements have allowed one to overcome the wavelength resolution limit.

In optics, the speed of light imposes strong instrumentation limits, which are relaxed for the speed of sound.
In acoustics, simultaneous measurements of the phase and amplitude are possible, for single frequencies or broadband signals. Hence no coherent source intertwines the recording and emitting tasks, i.e. they become separable \cite{muellerAcousticHolography1971}. In acoustic cloning, we fully reproduce a broadband scattering experiment. In contrast to the aforementioned classical holography, where the scattered wavefield is imaged on a small section of space over a limited frequency range, acoustic cloning images all multiple interactions between the clone and its surroundings in the volume and is broadband in time.

The inclusion of multiple broadband interactions allows us to exactly recreate the scattering of any wavefield incident on the hologram. The numerical representation of the scattering object permits digital manipulation of the object and taking physical measurements for altered or unphysical scatterers. Furthermore, we can completely synthesize numerical scattering experiments, which involve the physical scattering of the real object, without the object being present. The method is therefore not limited to being an imaging tool, but represents a research tool offering numerous applications in all domains of acoustic scattering, such as material characterization, active metamaterials, or even virtual acoustics applications.

Passive acoustic metasurfaces can be engineered to synthesize an acoustic hologram in transmission \cite{tian_acoustic_2017} and reflection \cite{zhu_fine_2018} at a specific frequency or even multiple discrete frequencies \cite{zhang_acoustic_2020,zhu_systematic_2021}. Combining transmission with reflection characteristics already becomes intractable for passive metasurfaces, even more so for broadband applications. Those limitations of passive acoustic metamaterials led to the  concept of active metamaterials \cite{ji_recent_2022}. Here, active control of the elastic parameters, such as density \cite{chen_active_2014}, stiffness, or modulus is used to tune the metamaterial to be able to cover a broader range of frequencies, for example through curved metasurfaces \cite{popa_active_2015,xiao-shuang_arbitrarily_2020}. In contrast to passive methods, active methods of scattering manipulation rely on the spatiotemporal control of distributed acoustic sources given a distribution of microphones. For acoustic cloning we rely completely on acoustic sources to control the pressure and particle velocities at the boundaries and inside our domain to prescribe the boundary conditions and clone the acoustic scatterer.

Similar to the initial holography proposal of \citet{gaborHolography194819711972}, combining the recorded phase information of light with imaging, here we draw on recent developments in data-driven Green's function retrieval and real-time experimentation to clone the response of arbitrary scatterers.
First, \citet{liClosedapertureUnboundedAcoustics2021} recently showed the retrieval of the free-space acoustic \gfs{} to characterize unknown objects from waveform data recorded in highly reverberant environments. Second, \citet{becker_broadband_2021} used full-wavefield extrapolators from finite-element simulations to conduct real-time, broadband acoustic cloaking and holography experiments for unknown incident fields.
Here, both methodological advances are combined to conceive a simple, general, two-step methodology for acoustic cloning.

The first step is to obtain the scattering \gf{} of an object under typical spatially constrained laboratory conditions. For this, the reverberations of the laboratory boundary need to be removed from the recorded data. State-of-the-art approaches for obtaining (reflection-free) radiation conditions are based on multidimensional deconvolution (MDD) \cite{liClosedapertureUnboundedAcoustics2021}, which is widely used in exploration seismology \cite{wapenaar_seismic_2011, amundsen_elimination_2001,amundsen_multidimensional_2000,vasmelModelindependentFinitedifferenceMethod2016}. Seismic data recorded in the subsurface can be postprocessed such that the scattered wavefield related to the overburden, or a reflecting top surface, are removed from the recorded data. One key feature of the MDD method is that it is purely data-driven: neither the medium above nor that below the receiver aperture, nor the source wavelet, need to be known to remove surface-related multiples \cite{amundsen_elimination_2001}. Similarly, for MDD in a laboratory environment, the material properties of a scattering object placed inside a closed receiver aperture do not need to be known a priori. This is consistent with typical laboratory characterization objectives. Furthermore, in order to remove the boundary-related scattering, it does not have to be characterized or known beforehand. 

The second step is broadband acoustic holography, employing the scattering \gf{} retrieved in step one. We implement the holography using immersive boundary conditions (IBCs), which were inspired by nonreflecting boundaries in numerical simulations \cite{tingExactBoundaryConditions1986, givoli_nonreflecting_1995}. However, IBCs are more general, as they enable embedding truncated modeling domains in larger, fixed, background media. Employing IBCs allow waves to propagate seamlessly between the truncated modeling domain and the background with arbitrary-order scattering interactions \cite{vanmanenExactWaveField2007, brogginiImmersiveBoundaryConditions2017}.
This is an important distinction of IBCs with acoustic wavefield synthesis, a related method. Acoustic wavefield synthesis is generally limited to synthesizing virtual source locations \cite{berkhoutAcousticControlWave1993,berkhoutArrayTechnologyAcoustic1997,devriesWaveFieldSynthesis1999}. In contrast, IBCs additionally modify the propagation of the existing field by interaction with virtual scatterers, a key requirement needed for acoustic cloning.
\citet{vasmelImmersiveExperimentationWave2013} described how IBCs can serve to embed physical wave propagation experiments in numerical simulations, thereby removing adverse reflections from boundaries of the experimental domain while fully capturing all interactions with the numerical background medium. This was subsequently demonstrated in one- and two-dimensional acoustic \cite{beckerImmersiveWavePropagation2018} as well as elastic \cite{thomsen_elastic_2019} experiments. Moreover, it has been shown numerically that IBCs can also be employed for cloaking and holography of broadband wavefields \cite{vanmanenBroadbandCloakingHolography2015}. Recently, cloaking and holography were verified experimentally in one and two dimensions \cite{borsingCloakingHolographyExperiments2019,becker_broadband_2021}.

For the demonstration of broadband holography in \cite{becker_broadband_2021} the reference scattering is derived from an initial design through manufacturing and the holographic scattering is derived from numerous full-waveform finite-element simulations. In contrast, for acoustic cloning we start with zero knowledge of an arbitrary scatterer and show how to replicate it exactly. We eliminate the need for precise manufacturing and costly simulations, which allows for direct on-demand cloning. The initial scatter is therefore the sole configuration needed to set up the hologram, such that not only the known properties but also the unknown properties of the scatterer are cloned. The difference in complexity with \cite{becker_broadband_2021} cannot be overstated: In finite-element simulations, the broad spatial and temporal bandwidth, as well as the radiation conditions, are obtained for free. In our experiments, the temporal and spatial bandwidth, as well as the appropriate absorbing conditions, have to be obtained from the band-limited reverberative scattering data. The success of our approach under such challenging and restrictive conditions confirms its generality and, indeed, justifies the very introduction of cloning as a concept.

The main body of the paper is structured as follows: We start by briefly reviewing the theory of scattering \gf{} retrieval, broadband cloaking and holography. Then we explain in detail how these approaches can be combined to clone an object acoustically and illustrate the resulting methodology with two-dimensional (2D) real-data examples. We further show how the clone can be translated spatially and how gain and transparency not present in the original scatterer can be included in the cloning, effectively making it an augmented clone. Finally, we discuss the characteristics that make our approach an efficient and general methodology.

\section{Theory}
\label{sec:theory}

\subsection{\gf{} retrieval using MDD}
\label{sec:mdd}

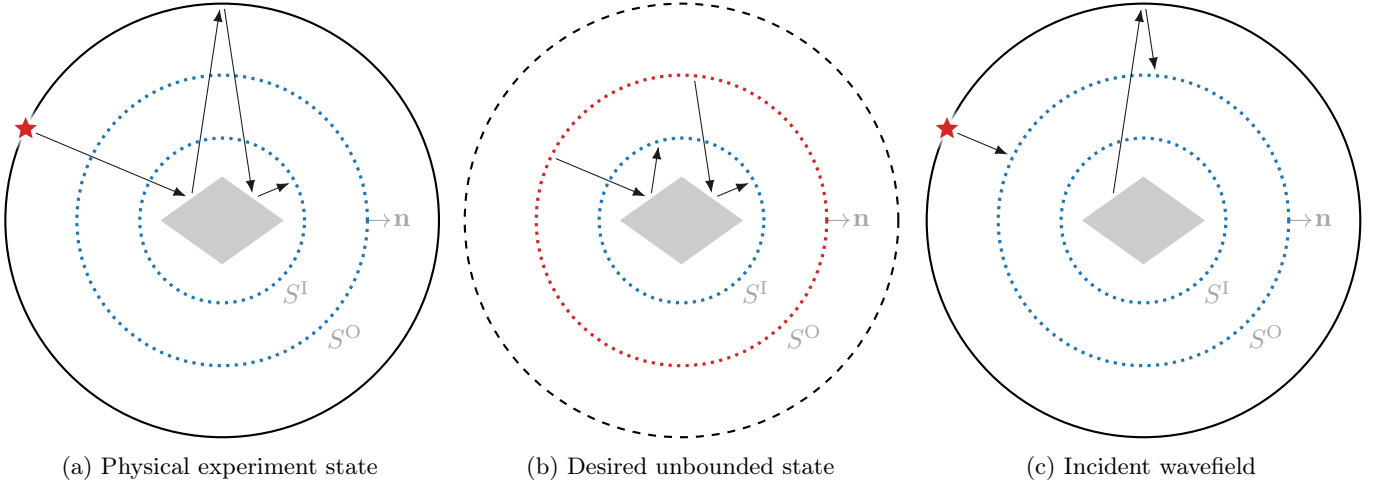
\begin{figure*}[t]
	\textbf{Step 1: Green's Function Retrieval}\\ \vspace{\baselineskip}
	\begin{subfigure}[c]{0.32\textwidth}
	\begin{tikzpicture}[]
		\def\rdomain{\textwidth/2}
\def\rrec{0.67*\rdomain}
\def\rsrc{0.38*\rdomain}

\def\scatL{2*0.75*\rsrc} \def\scatW{\scatL/1.4}

\def\phi{65}

\coordinate (source) at (90+\phi:\rdomain);

\coordinate (scatTop) at (0,\scatW/2);
\coordinate (scatLeft) at (-\scatL/2,0);
\coordinate (scatRight) at (\scatL/2,0);

\coordinate (scat1) at ($(scatTop)!0.5!(scatLeft)$);
\coordinate (ref1) at (90:\rdomain);
\coordinate (scat2) at ($(scatTop)!0.5!(scatRight)$);
\coordinate (ref2) at (90-\phi:\rdomain);

\coordinate (ref1Left) at ($(ref1)-(0.02,0)$);
\coordinate (ref1Right) at ($(ref1)+(0.02,0)$);

\draw[name path=in1, draw=none] (source)   -- (scat1);
\draw[name path=out1, draw=none] (scat1) -- ($(ref1)-(0.02,0)$);
\draw[name path=in2, draw=none] ($(ref1)+(0.02,0)$) -- (scat2);
\draw[name path=out2, draw=none] (scat2) -- (ref2);

\node[circle,label=below right:\textcolor{colAnotations}{$\si$},minimum size=\rsrc*2-8pt] {};
\node[circle,label=below right:\textcolor{colAnotations}{$\so$},minimum size=\rrec*2-8pt] {};

\node[inner sep=1pt,colAnotations] at (0:\rrec+13pt) (normal) {$\mathbf{n}$};
\draw[->,colAnotations] (0:\rrec)  -- (normal); 		\node[scat] (d) at (0,0) {};
		
\draw[black,thick] circle [radius=\rdomain];
		\draw[rec,name path=outer] circle [radius=\rrec];
		\draw[rec,name path=inner] circle [radius=\rsrc];

\fill[white, path fading=circle with fuzzy edge 50 percent] (source) circle [radius=3mm];
        \node[sourceStar] (star) at (source) {};
        
\draw[ray] (star)   -- (scat1);
		\draw[ray] (scat1) -- (ref1Left);
		\draw[ray] (ref1Right) -- (scat2);
		
		\path [name intersections={of=inner and out2,by=int}];
		\draw[ray] (scat2) -- (int);
	   
	\end{tikzpicture}
	\caption{Physical experiment state}
	\label{fig:mddPhysical}
\end{subfigure}
\hfill
\begin{subfigure}[c]{0.32\textwidth}
	\begin{tikzpicture}[]
		\def\rdomain{\textwidth/2}
\def\rrec{0.67*\rdomain}
\def\rsrc{0.38*\rdomain}

\def\scatL{2*0.75*\rsrc} \def\scatW{\scatL/1.4}

\def\phi{65}

\coordinate (source) at (90+\phi:\rdomain);

\coordinate (scatTop) at (0,\scatW/2);
\coordinate (scatLeft) at (-\scatL/2,0);
\coordinate (scatRight) at (\scatL/2,0);

\coordinate (scat1) at ($(scatTop)!0.5!(scatLeft)$);
\coordinate (ref1) at (90:\rdomain);
\coordinate (scat2) at ($(scatTop)!0.5!(scatRight)$);
\coordinate (ref2) at (90-\phi:\rdomain);

\coordinate (ref1Left) at ($(ref1)-(0.02,0)$);
\coordinate (ref1Right) at ($(ref1)+(0.02,0)$);

\draw[name path=in1, draw=none] (source)   -- (scat1);
\draw[name path=out1, draw=none] (scat1) -- ($(ref1)-(0.02,0)$);
\draw[name path=in2, draw=none] ($(ref1)+(0.02,0)$) -- (scat2);
\draw[name path=out2, draw=none] (scat2) -- (ref2);

\node[circle,label=below right:\textcolor{colAnotations}{$\si$},minimum size=\rsrc*2-8pt] {};
\node[circle,label=below right:\textcolor{colAnotations}{$\so$},minimum size=\rrec*2-8pt] {};

\node[inner sep=1pt,colAnotations] at (0:\rrec+13pt) (normal) {$\mathbf{n}$};
\draw[->,colAnotations] (0:\rrec)  -- (normal); 		\node[scat] (d) at (0,0) {};
		
\draw[black,dashed,thick] circle [radius=\rdomain];
		\draw[src,name path=outer] circle [radius=\rrec];
		\draw[rec,name path=inner] circle [radius=\rsrc];
		
\path [name intersections={of=outer and in1,by=int}];
		\draw[ray] (int)   -- (scat1);
		
		\path [name intersections={of=inner and out1,by=int}];
		\draw[ray] (scat1) -- (int);
		
		\path [name intersections={of=outer and in2,by=int}];
		\draw[ray] (int) -- (scat2);
		
		\path [name intersections={of=inner and out2,by=int}];
		\draw[ray] (scat2) -- (int);
		
	\end{tikzpicture}
	\caption{Desired unbounded state}
	\label{fig:mddDesired}
\end{subfigure}
\hfill
	\begin{subfigure}[c]{0.32\textwidth}
	\begin{tikzpicture}[]
		\def\rdomain{\textwidth/2}
\def\rrec{0.67*\rdomain}
\def\rsrc{0.38*\rdomain}

\def\scatL{2*0.75*\rsrc} \def\scatW{\scatL/1.4}

\def\phi{65}

\coordinate (source) at (90+\phi:\rdomain);

\coordinate (scatTop) at (0,\scatW/2);
\coordinate (scatLeft) at (-\scatL/2,0);
\coordinate (scatRight) at (\scatL/2,0);

\coordinate (scat1) at ($(scatTop)!0.5!(scatLeft)$);
\coordinate (ref1) at (90:\rdomain);
\coordinate (scat2) at ($(scatTop)!0.5!(scatRight)$);
\coordinate (ref2) at (90-\phi:\rdomain);

\coordinate (ref1Left) at ($(ref1)-(0.02,0)$);
\coordinate (ref1Right) at ($(ref1)+(0.02,0)$);

\draw[name path=in1, draw=none] (source)   -- (scat1);
\draw[name path=out1, draw=none] (scat1) -- ($(ref1)-(0.02,0)$);
\draw[name path=in2, draw=none] ($(ref1)+(0.02,0)$) -- (scat2);
\draw[name path=out2, draw=none] (scat2) -- (ref2);

\node[circle,label=below right:\textcolor{colAnotations}{$\si$},minimum size=\rsrc*2-8pt] {};
\node[circle,label=below right:\textcolor{colAnotations}{$\so$},minimum size=\rrec*2-8pt] {};

\node[inner sep=1pt,colAnotations] at (0:\rrec+13pt) (normal) {$\mathbf{n}$};
\draw[->,colAnotations] (0:\rrec)  -- (normal); 		\node[scat] (d) at (0,0) {};
		
\draw[black,thick] circle [radius=\rdomain];
		\draw[rec,name path=outer] circle [radius=\rrec];
		\draw[rec,name path=inner] circle [radius=\rsrc];

\fill[white, path fading=circle with fuzzy edge 50 percent] (source) circle [radius=3mm];
        \node[sourceStar] (star) at (source) {};
          
\draw[ray] (scat1) -- (ref1Left);
		
		\path [name intersections={of=outer and in1,by=int}];
		\draw[ray] (star)   -- (int);
		
		\path [name intersections={of=outer and in2,by=int}];
		\draw[ray] (ref1Right) -- (int);
		
	\end{tikzpicture}
	\caption{Incident wavefield}
	\label{fig:mddIncident}
\end{subfigure} 	\caption{Configuration for scattering \gfs{} retrieval using multidimensional deconvolution. The sound transparent outer surface $\so$ and inner surface $\si$ are represented by dotted circles. A recording surface is indicated by blue dots and an emitting surface by red dots. In the physical experiment state \subref{fig:mddPhysical}, a possible ray path for an external excitation (outside of $\so$) is shown. It undergoes multiple reflections on the rigid surface of the scatterer and waveguide boundary. The desired experiment state \subref{fig:mddDesired} has radiation boundary conditions (dashed black circle) with the unknown \gfs{} connecting the arrivals on $\so$ \subref{fig:mddIncident} to their destination on $\si$. The multidimensional convolution of the incident field on $\so$ \subref{fig:mddIncident} and \gf{} \subref{fig:mddDesired} corresponds to the physical experiment \subref{fig:mddPhysical} such that we can roughly represent \cref{eq:mdc} graphically as $\subref{fig:mddPhysical}=\subref{fig:mddDesired}\ast \subref{fig:mddIncident}$.}
	\label{fig:MDDconfig}
\end{figure*}

Let us consider the setup shown in \cref{fig:MDDconfig}, similar to the setup used by \citet{liClosedapertureUnboundedAcoustics2021} for scattering \gf{} retrieval using multidimensional deconvolution (MDD). It is characterized by the two surfaces $S^{\mathrm{Inner}}\equiv \si$ and $S^{\mathrm{Outer}}\equiv \so$. The experimental domain $V$ is thus subdivided into ${V^\mathrm{I} \subset V^\mathrm{O} \subset V}$ such that $\partial V^\mathrm{I} = \si$ and $\partial V^\mathrm{O} = \so$ as in \cref{fig:MDDconfig}.

We consider two different states: a physical experiment with reflecting boundary conditions, \cref{fig:mddPhysical}, and a desired experiment with radiation boundary conditions, \cref{fig:mddDesired}. Using reciprocity theory \cite{fokkemaSeismicApplicationsAcoustic1993, hoopHandbookRadiationScattering1995} we derive two expressions relating the pressure on the inner surface $p^\mathrm{I}\equiv p(\xin,t)$, to the ingoing pressure and ingoing normal particle velocity on the outer surface $\{p,v\}^\mathrm{in}(\so)$. The relations contain a set of monopolar and dipolar scattering \gfs{} $\gpq$ and $\gpf$ \cite{liClosedapertureUnboundedAcoustics2021,ravasiSeismicInterferometryMultidimensional2015,wapenaar_seismic_2011}
\begin{subequations}
	\begin{alignat}{2}
			p(\xin,t) &=-&2\oint_{\so}   \gpq(\xin,t\mid \xout)* v^{\mathrm{in}}(\xout,t) \,dS,\\ 
			p(\xin,t) &=&2\oint_{\so}   \gpf(\xin,t\mid \xout)*p^{\mathrm{in}}(\mathbf{\xout},t)  \,dS,
		\end{alignat}
		\label{eq:mdc}\end{subequations}
with $*$ denoting the temporal convolution operator and ${G^{p\mid q}(\xin,t \mid \xout)}$ the \gf{} of the medium evaluating to a pressure at $\xin \in \si$ due to a point source of volume injection rate (monopole) at $\xout \in \so$.
We only consider normal components of vector fields on surfaces, so that we can lighten the notation with $v \equiv v_n \equiv \mathbf{n} \cdot \mathbf{v}$ and $\mathbf{n}$ the outward-pointing normal on a surface $S$. Additionally  $\gpf\equiv G^{p\mid f_n} = \mathbf{n} \cdot \mathbf{G}^{p\mid f}$. This notation will be used henceforth. Furthermore we will also use $\{G^{v\mid f} \equiv G^{v_n\mid f_n}\}(\xin,t \mid \xout)$ representing the particle velocity normal to the evaluation surface $\si$ due a point source of volume force (dipole) normal to the integration surface $\so$ \cite{hoopHandbookRadiationScattering1995}.
Equations~(\ref{eq:mdc}) describe a multidimensional convolution (MDC) of the incident wavefield (\cref{fig:mddIncident}) and the \gfs{} of the desired unbounded state (\cref{fig:mddDesired}). We can solve for the unknown \gfs{} by using recorded pressure $p(\xin,t)$ and the ingoing wavefield $\{p,v\}^\mathrm{in}(\xout,t)$ from a physical experiment \cite{liClosedapertureUnboundedAcoustics2021, sonnelandDeghostingUsingVertical1986}. A detailed description of the wavefield separation process is given in \Cref{app:separation}.

Equations of the form \eqref{eq:mdc} are Fredholm integrals of the first kind, the inversion of which is highly ill-posed if wavefields $\{p^\mathrm{I},p_\mathrm{in}^\mathrm{O}\}$ or $\{p^\mathrm{I},v_\mathrm{in}^\mathrm{O}\}$ for only a single exterior source are considered \cite{liClosedapertureUnboundedAcoustics2021, aster_parameter_2013}. Hence, one needs to excite multiple sources exterior to $\so$ while recording the wavefield on both surfaces. These wavefields can then be separated into incident and outgoing, with which we can compose a system of equations by considering multiple realizations of \cref{eq:mdc} simultaneously \cite{liClosedapertureUnboundedAcoustics2021, wapenaar_passive_2008}. This system of equations can then be inverted in a least-squares sense using multidimensional deconvolution \cite{liClosedapertureUnboundedAcoustics2021, amundsen_elimination_2001, van_der_neut_data_2012}.

The sought-after \gfs{} in \cref{eq:mdc} involve the impulse responses of a desired medium that only includes the physical scattering of the real medium inside $\si$, as illustrated in \cref{fig:mddDesired} \cite{liClosedapertureUnboundedAcoustics2021}. Scatterers exterior to $\so$, such as the reflecting boundary of the experimental domain in \cref{fig:mddPhysical}, do not exist in the desired medium. This highlights the strength of using MDD in laboratory experiments for the removal of undesired scattering effects from the physical boundary. Hereby, recorded data are postprocessed such that only waveforms related to the interior scatterers are retrieved. As noted by \hbox{\citet{liClosedapertureUnboundedAcoustics2021}}, the illuminating sources outside $\so$ do not need to be densely or regularly spaced and can be placed without consideration of the Nyquist criterion. They can even be a noise wavefield \cite{sternini_bistatic_2020}. Crucially, the source spectra and temporal signatures do not need to be known in the physical experiment. In fact, any source characteristics, including radiation patterns, will be removed by MDD. Finally, because it is based on a convolutional representation, MDD works in the presence of loss and for different types of wave physics inside $\si$ (e.g. elastic and viscoelastic) \cite{wapenaar_seismic_2011}. A detailed flowchart for \gf{} retrieval using MDD is shown in \Cref{app:mdd}.

In the absence of any scatterer within $\si$, we speak of the homogeneous state. The experimental domain $V$ then constitutes a single homogeneous medium like air. If a scatterer is placed within $\si$, we denote this as the heterogeneous state. The experimental domain then contains different media: air and a scatterer. The difference in the wavefield propagating in both states is then solely due to the presence of the scatterer. It can then be easily isolated by subtracting the homogeneous from the heterogeneous wavefield. We then denote this as the scattered wavefield.

For cloning, we are interested only in the scattered part of the \gfs{}.
We can therefore subtract the results from a second MDD experiment for a homogeneous interior from the heterogeneous \gfs{}, yielding $G_\mathrm{scat} = G_\mathrm{het}-G_\mathrm{hom}$. This step allows us to isolate the \gfs{} of the scatterer only, since the repeatability of the experiment ensures that the direct arrival is the same in the homogeneous and heterogeneous \gfs{}.

\subsection{Real-time broadband Holography}

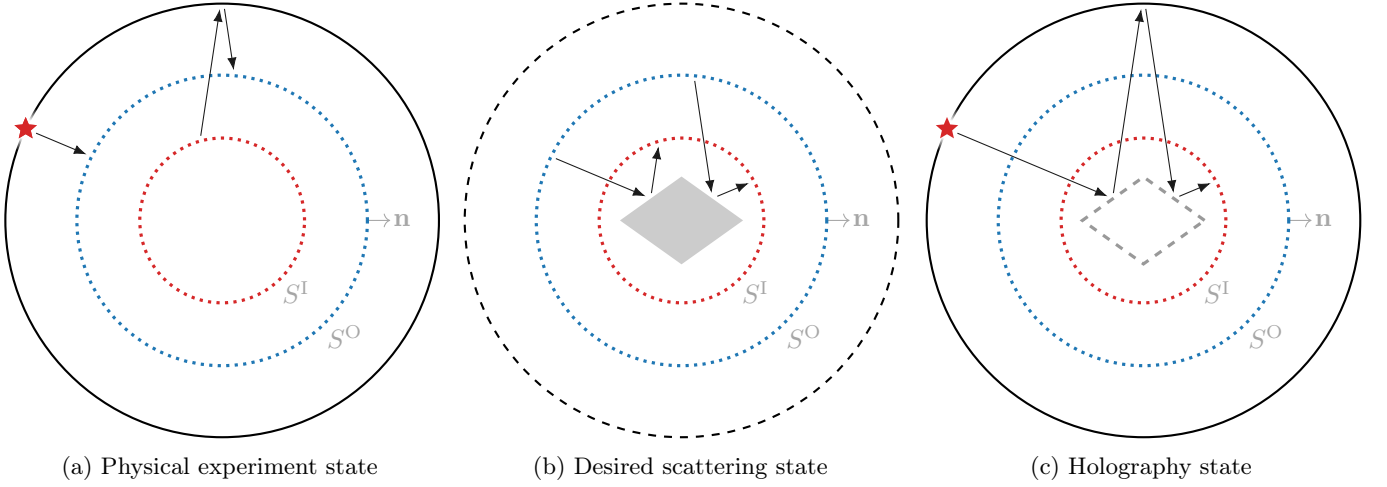
\begin{figure*}[t]
	\textbf{Step 2: Real-Time Holography}\\ \vspace{\baselineskip}
		\begin{subfigure}[c]{0.32\textwidth}
	\begin{tikzpicture}[]
		\def\noScat{1}
		\def\rdomain{\textwidth/2}
\def\rrec{0.67*\rdomain}
\def\rsrc{0.38*\rdomain}

\def\scatL{2*0.75*\rsrc} \def\scatW{\scatL/1.4}

\def\phi{65}

\coordinate (source) at (90+\phi:\rdomain);

\coordinate (scatTop) at (0,\scatW/2);
\coordinate (scatLeft) at (-\scatL/2,0);
\coordinate (scatRight) at (\scatL/2,0);

\coordinate (scat1) at ($(scatTop)!0.5!(scatLeft)$);
\coordinate (ref1) at (90:\rdomain);
\coordinate (scat2) at ($(scatTop)!0.5!(scatRight)$);
\coordinate (ref2) at (90-\phi:\rdomain);

\coordinate (ref1Left) at ($(ref1)-(0.02,0)$);
\coordinate (ref1Right) at ($(ref1)+(0.02,0)$);

\draw[name path=in1, draw=none] (source)   -- (scat1);
\draw[name path=out1, draw=none] (scat1) -- ($(ref1)-(0.02,0)$);
\draw[name path=in2, draw=none] ($(ref1)+(0.02,0)$) -- (scat2);
\draw[name path=out2, draw=none] (scat2) -- (ref2);

\node[circle,label=below right:\textcolor{colAnotations}{$\si$},minimum size=\rsrc*2-8pt] {};
\node[circle,label=below right:\textcolor{colAnotations}{$\so$},minimum size=\rrec*2-8pt] {};

\node[inner sep=1pt,colAnotations] at (0:\rrec+13pt) (normal) {$\mathbf{n}$};
\draw[->,colAnotations] (0:\rrec)  -- (normal); 		
\draw[black,thick] circle [radius=\rdomain];
		\draw[rec,name path=outer] circle [radius=\rrec];
		\draw[src,name path=inner] circle [radius=\rsrc];

\fill[white, path fading=circle with fuzzy edge 50 percent] (source) circle [radius=3mm];
        \node[sourceStar] (star) at (source) {};
        
\path [name intersections={of=inner and out1,by=int}];
		\draw[ray] (int) -- (ref1Left);
		
		\path [name intersections={of=outer and in1,by=int}];
		\draw[ray] (star)   -- (int);
		
		\path [name intersections={of=outer and in2,by=int}];
		\draw[ray] (ref1Right) -- (int);
		
\node[sourceStar] at (source) {};
	\end{tikzpicture}
	\caption{Physical experiment state}
	\label{fig:holographyIncident}
\end{subfigure}
\hfill
\begin{subfigure}[c]{0.32\textwidth}
	\begin{tikzpicture}[]
		\def\rdomain{\textwidth/2}
\def\rrec{0.67*\rdomain}
\def\rsrc{0.38*\rdomain}

\def\scatL{2*0.75*\rsrc} \def\scatW{\scatL/1.4}

\def\phi{65}

\coordinate (source) at (90+\phi:\rdomain);

\coordinate (scatTop) at (0,\scatW/2);
\coordinate (scatLeft) at (-\scatL/2,0);
\coordinate (scatRight) at (\scatL/2,0);

\coordinate (scat1) at ($(scatTop)!0.5!(scatLeft)$);
\coordinate (ref1) at (90:\rdomain);
\coordinate (scat2) at ($(scatTop)!0.5!(scatRight)$);
\coordinate (ref2) at (90-\phi:\rdomain);

\coordinate (ref1Left) at ($(ref1)-(0.02,0)$);
\coordinate (ref1Right) at ($(ref1)+(0.02,0)$);

\draw[name path=in1, draw=none] (source)   -- (scat1);
\draw[name path=out1, draw=none] (scat1) -- ($(ref1)-(0.02,0)$);
\draw[name path=in2, draw=none] ($(ref1)+(0.02,0)$) -- (scat2);
\draw[name path=out2, draw=none] (scat2) -- (ref2);

\node[circle,label=below right:\textcolor{colAnotations}{$\si$},minimum size=\rsrc*2-8pt] {};
\node[circle,label=below right:\textcolor{colAnotations}{$\so$},minimum size=\rrec*2-8pt] {};

\node[inner sep=1pt,colAnotations] at (0:\rrec+13pt) (normal) {$\mathbf{n}$};
\draw[->,colAnotations] (0:\rrec)  -- (normal); 		\node[scat] (d) at (0,0) {};
		
\draw[black,dashed,thick] circle [radius=\rdomain];
		\draw[rec,name path=outer] circle [radius=\rrec];
		\draw[src,name path=inner] circle [radius=\rsrc];
		
\path [name intersections={of=outer and in1,by=int}];
		\draw[ray] (int)   -- (scat1);
		
		\path [name intersections={of=inner and out1,by=int}];
		\draw[ray] (scat1) -- (int);
		
		\path [name intersections={of=outer and in2,by=int}];
		\draw[ray] (int) -- (scat2);
		
		\path [name intersections={of=inner and out2,by=int}];
		\draw[ray] (scat2) -- (int);
		
	\end{tikzpicture}
	\caption{Desired scattering state}
	\label{fig:holographyDesired}
\end{subfigure}
\hfill
\begin{subfigure}[c]{0.32\textwidth}
	\begin{tikzpicture}[]
		\def\rdomain{\textwidth/2}
\def\rrec{0.67*\rdomain}
\def\rsrc{0.38*\rdomain}

\def\scatL{2*0.75*\rsrc} \def\scatW{\scatL/1.4}

\def\phi{65}

\coordinate (source) at (90+\phi:\rdomain);

\coordinate (scatTop) at (0,\scatW/2);
\coordinate (scatLeft) at (-\scatL/2,0);
\coordinate (scatRight) at (\scatL/2,0);

\coordinate (scat1) at ($(scatTop)!0.5!(scatLeft)$);
\coordinate (ref1) at (90:\rdomain);
\coordinate (scat2) at ($(scatTop)!0.5!(scatRight)$);
\coordinate (ref2) at (90-\phi:\rdomain);

\coordinate (ref1Left) at ($(ref1)-(0.02,0)$);
\coordinate (ref1Right) at ($(ref1)+(0.02,0)$);

\draw[name path=in1, draw=none] (source)   -- (scat1);
\draw[name path=out1, draw=none] (scat1) -- ($(ref1)-(0.02,0)$);
\draw[name path=in2, draw=none] ($(ref1)+(0.02,0)$) -- (scat2);
\draw[name path=out2, draw=none] (scat2) -- (ref2);

\node[circle,label=below right:\textcolor{colAnotations}{$\si$},minimum size=\rsrc*2-8pt] {};
\node[circle,label=below right:\textcolor{colAnotations}{$\so$},minimum size=\rrec*2-8pt] {};

\node[inner sep=1pt,colAnotations] at (0:\rrec+13pt) (normal) {$\mathbf{n}$};
\draw[->,colAnotations] (0:\rrec)  -- (normal); 		\node[diamond, dashed, draw = gray!80, very thick, minimum width = \scatL, minimum height = \scatW] (d) at (0,0) {};
		
\draw[black,thick] circle [radius=\rdomain];
		\draw[rec,name path=outer] circle [radius=\rrec];
		\draw[src,name path=inner] circle [radius=\rsrc];

\fill[white, path fading=circle with fuzzy edge 50 percent] (source) circle [radius=3mm];
        \node[sourceStar] (star) at (source) {};
        
\draw[ray] (star)   -- (scat1);
		\draw[ray] (scat1) -- (ref1Left);
		\draw[ray] (ref1Right) -- (scat2);
		
		\path [name intersections={of=inner and out2,by=int}];
		\draw[ray] (scat2) -- (int);
        
	\end{tikzpicture}
	\caption{Holography state}
	\label{fig:holographyAugmented}
\end{subfigure} 	\caption{Real-time broadband holography configuration. Dotted circles represent outer recording surface $\so$ (blue) and inner emitting surface $\si$ (red) respectively. The ingoing field on $\so$ in \subref{fig:holographyIncident} is extrapolated using the scattering \gfs{} of the object to be recreated in \subref{fig:holographyDesired}. The holographic field \subref{fig:holographyAugmented} is then any physical field  \subref{fig:holographyIncident} propagating from $\so$ to $\si$, where it scatters on the immersive boundary conditions (IBCs). The IBCs represent the scatterer through \gfs{} \subref{fig:holographyDesired} as the numerical extrapolator from $\so$ to $\si$.}
	\label{fig:holgraphyConfig}
\end{figure*}

Similar to the MDD configuration in \cref{fig:MDDconfig}, the configuration for broadband cloaking and holography considered by \citet{becker_broadband_2021} is shown in \cref{fig:holgraphyConfig}. Here, the inner surface $\si$ is of emitting type while $\so$ is again of recording type. The inner surface $\si$ may coincide with the surface of a scatterer or with a transparent surface. The acoustic pressure in the homogeneous state, no scatterer present, is denoted by $p_\mathrm{hom}$. This state can be augmented by applying appropriate boundary conditions on the emitting surface $\si$ resulting in the superposition state \cite{vanmanenBroadbandCloakingHolography2015},
\begin{equation}
	p_\mathrm{het}(\mathbf{x},t) = p_\mathrm{hom}(\mathbf{x},t) + p_\mathrm{scat}(\mathbf{x},t)
	\quad \forall \mathbf{x} \in V^\mathrm{lab},
	\label{eq:p_aug}
\end{equation}
with $V^\mathrm{lab} = V\setminus V^\mathrm{I}$. Using immersive boundary conditions (IBCs) on $\si$, we thus aim to synthesize the wavefield of an arbitrary virtual scatterer $p_\mathrm{scat}$ to obtain the desired wavefield $p_\mathrm{het}$. A surface integral representation for $p_\mathrm{scat}(\si)$ can be found by considering reciprocity between the homogeneous state, \cref{fig:holographyIncident}, and the heterogeneous state, \cref{fig:holographyDesired} \cite{fokkemaSeismicApplicationsAcoustic1993, hoopHandbookRadiationScattering1995}. This gives the following expression for cloaking and holography configurations involving internal surfaces \cite{vanmanenBroadbandCloakingHolography2015},
\begin{flalign}
	p_\mathrm{scat}(\mathbf{x},t) = \oint_{\si}[ &
	G_\mathrm{hom}^{p \mid q}(\mathbf{x},t\mid \xin) \ast v_\mathrm{scat}(\xin,t) \nonumber    \\
	+ &G_\mathrm{hom}^{p\mid f}(\mathbf{x},t \mid \xin)  \ast p_\mathrm{scat}(\xin,t)] dS,
	\label{eq:p_ibc}
\end{flalign}
for the domain $\mathbf{x} \in V^\mathrm{lab}$ and with $\{p,v\}_\mathrm{scat}=\{p,v\}_\mathrm{het}-\{p,v\}_\mathrm{hom}$ being the wavefield due to the presence of the scatterer only.
\label{sec:holography}
The surface integral over $\si$ in \cref{eq:p_ibc} can be understood as summing the contributions of densely spaced source distributions, of type monopole $q$ and dipole $f$, driven with the time-varying strengths $\{p,v\}_\mathrm{scat}(\mathbf{x}^\mathrm{I},t)$. Note that, due to reciprocity, the particle velocity $v$  drives the monopole \gf{} $G_\mathrm{hom}^{p\mid q}$ and the pressure $p$ the dipole \gf{} $G_\mathrm{hom}^{p\mid f}$. Despite the apparent simplicity of \cref{eq:p_ibc}, its implementation in real-time experiments poses significant challenges, as the pressure and the particle velocity need to be known on $\si$ prior to the wavefield arriving there \cite{becker_broadband_2021}.

\begin{figure*}[t]
	\begin{tikzpicture}
\def\width{17cm}
	\def\domain{4.5pt}
	
	\def\scaleInset{0.9}
	\def\height{1.6cm}
	\def\depth{-2cm}
	\def\xs{\width/2*\scaleInset}
	\def\sDist{0.6cm}

\node at (-\width/2,0) (left) {};
	\node at (\width/2,0) (right) {};

\draw (left) -- (right);
	\draw [decoration={text along path,text={physical},text align=center,raise=\domain+0.6pt},decorate] (left) -- (right);
	\draw [decoration={text along path,text={numerical},text align=center,raise=\domain-\baselineskip*1.25},decorate] (left) -- (right);

\draw[dashed,\colrec,very thick] (-\xs,0) -- (-\xs,\height);

	\node[anchor=south] at (-\xs,\height) {$x^\mathrm{o}$};
	
	{\scriptsize
	\node[recShape,scale=0.8] at (-\xs-\sDist,\height*0.55) {};
	\node[recShape,scale=0.8] at (-\xs+\sDist,\height*0.55) {};
	}
	
\draw[dashed,\colsrc,very thick] (\xs,0) -- (\xs,\height);
	 \node [anchor=south] at (\xs,\height) {$x^\mathrm{i}$};

	 \node[sourceStar,scale=1.2] at (\xs-\sDist,\height*0.55) {};
	 \node[sourceStar,scale=1.2] at (\xs+\sDist,\height*0.55) {};
	 
\draw[-Latex,decoration={snake, segment length=\width*0.7/3*1.12, amplitude=1.5mm},
	 decorate,
	 postaction={decorate,decoration={text along path,text={direct arrival},text align=center,raise=2.5pt}},
	 ] (-\width/2*0.7,\height/2) --  (\width/2*0.7,\height/2);

\node[rectangle,draw] (gf) at (-6cm,\depth) {$
	 	*
	 \begin{bmatrix}
	 	   \\
	 	  G(x^\mathrm{i} \mid x^\mathrm{o}) \\
	 	 \\
	 \end{bmatrix}
	$};

	\node[circle,draw] (buffer) at (0cm,\depth) {+};
	
	\node[isosceles triangle,isosceles triangle apex angle=60,draw] (stageOut)at (6cm,\depth) {$t_s$};
	
\node[circle,draw,fill=white,opacity=1] (output) at (\xs,0) {+};

\draw[-Latex] (-\xs,0)  |- (gf.west)  node [pos=0.16,right,align=center ] {$
		\begin{bmatrix}
			p(x^\mathrm{o}_+)\\
			p(x^\mathrm{o}_-)
		\end{bmatrix}(t_s)
	$} ;

\draw[-Latex] (gf.east)  |- (buffer.west)  node [pos=0.75,below,align=center ] {$
	\begin{bmatrix}
		p(x^\mathrm{i})\\
		v(x^\mathrm{i})
	\end{bmatrix}(t_s,\ldots, t_s+T)
	$};
	\draw[-Latex] (buffer.east) -- ++(0.5cm,0) coordinate (branch) |- ++(0,-0.8cm) -| (buffer.south) {};
	
	\draw[-Latex] (branch) -- (stageOut.west) node [pos=0.5,below,align=center ] {$
		\begin{bmatrix}
			p(x^\mathrm{i})\\
			v(x^\mathrm{i})
		\end{bmatrix}(0,\ldots, t_s+T)
	$};
    \fill (branch) circle (2pt);
	
	\draw[-Latex] (stageOut.east) -| (output.south) node [pos=0.9,left,align=center ] {$
		\begin{bmatrix}
			p(x^\mathrm{i}_+)\\
			p(x^\mathrm{i}_-)
		\end{bmatrix}(t_s)
		$};
	
\end{tikzpicture} 	\caption{Illustration of wave propagation in the physical domain and simultaneous numerical extrapolation from $x^o\in \so$ to $x^i\in \si$. From two finite-difference pressure recordings (blue triangles) we calculate two finite-difference pressures to apply on the loudspeakers (red stars).
Since in the physical domain waves can travel any $x^o \rightarrow x^i$ the same extrapolation has to be done in the numerical domain. Hence there are $N^\mathrm{rec} N^\mathrm{emt} = 114 \times 18 = 2052$ extrapolations at each time step. 
}
	\label{fig:setupIbc}
\end{figure*}
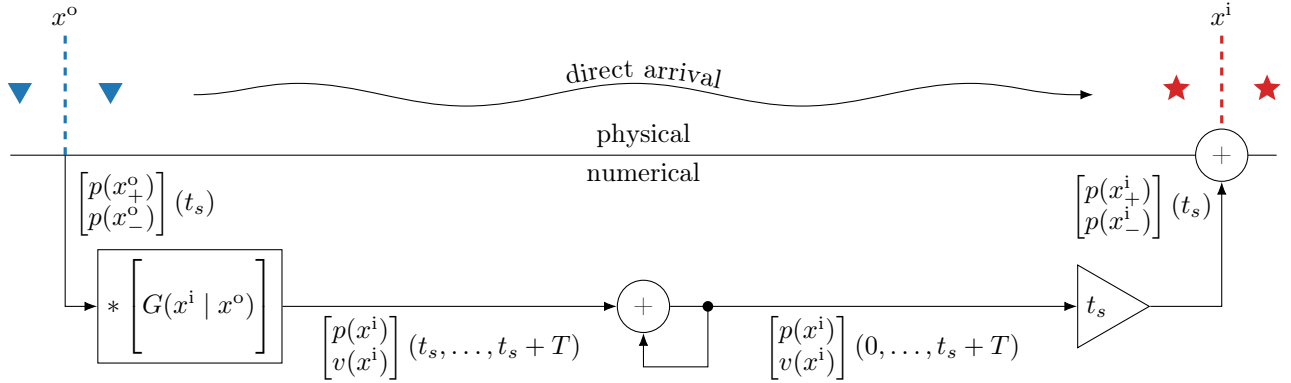

Furthermore, $\{p,v\}_\mathrm{scat}$ in \cref{eq:p_ibc} depend on the heterogeneous field they are creating and thus have to be constructed recursively. For this, we resort to the outer surface $\so$ enclosing $\si$. On $\so$ the wavefield $\{p,v\}$ is recorded and extrapolated to $\si$ using Kirchhoff-Helmholtz integrals. We can then use the wavefield propagation time $\{p,v\}(\so) \mapsto \{p,v\}(\si)$ to predict the field that will be scattered within $\si$ by the virtual scatterer $G_\mathrm{scat}$ as \cite{becker_broadband_2021}
\begin{subequations}
		\begin{equation}
		\begin{aligned}
			p_\mathrm{scat}(\xin,t) = \oint_{\so}[  &G_\mathrm{scat}^{p\mid q}(\xin,t\mid \xout) \ast v_\mathrm{het}(\xout,t) \\
			+ &G_\mathrm{scat}^{p\mid f}(\xin,t \mid \xout) \ast p_\mathrm{het}(\xout,t)]dS,
		\end{aligned}
		\label{eq:extrap_p}
	\end{equation}
	\begin{equation}
		\begin{aligned}
			v_\mathrm{scat}(\xin,t)= \oint_{\so}[  &G_\mathrm{scat}^{v\mid q}(\xin,t\mid \xout) \ast v_\mathrm{het}(\xout,t) \\
					+ &G_\mathrm{scat}^{v\mid f}(\xin,t\mid \xout) \ast p_\mathrm{het}(\xout,t)]dS,
		\end{aligned}
		\label{eq:extrap_v2}
	\end{equation}
	\label{eq:extrap}\end{subequations}
with $G_\mathrm{scat} = G_\mathrm{het} - G_\mathrm{hom}$. This subtraction isolates the scattering \gf{} since $G_\mathrm{hom}$ contains only the direct waves propagating from $\so$ to $\si$ and $G_\mathrm{het}$ additionally contains the scattering contribution of the inhomogeneity (scatterer).
Therefore, the right-hand sides of \cref{eq:extrap} extrapolate the field from $\so$ to the scattered field on $\si$. The wavefield $\{p,v\}(\so)$ physically propagates from $\so$ to $\si$ while simultaneously being extrapolated in the numerical domain through \cref{eq:extrap} i.e., the physical and numerical domain overlap in the region between $\so$ and $\si$. This is illustrated for one channel connecting $\mathbf{x}^o\in \so$ to $\mathbf{x}^i\in \si$ in \cref{fig:setupIbc}.

From here we can easily understand immersive boundary conditions. Given a primary wavefield $\{p,v\}(\so,t)$, using \cref{eq:extrap} we can predict the field $\{p,v\}_\mathrm{scat}(\si,t)$. Using this field as source signatures for monopole and dipole sources on $\si$ as described by \cref{eq:p_ibc} will result in the exact construction of the scattered wavefield within $V^\mathrm{lab}$. From any point $x\in V^\mathrm{lab}$, the illusion of a scatterer existing within the subdomain $V^\mathrm{I}$ is created, since any wavefield probing $V^\mathrm{I}$ from $V^\mathrm{ext} = V \setminus V^\mathrm{O}$ will scatter on the hologram characterized by $G_\mathrm{scat}$. A detailed flowchart for real-time holography is included in \cref{fig:flowholography} of \Cref{app:mdd}.

Note that the simplicity of \cref{eq:p_aug} hides the strengths and complexities of IBCs by giving the impression of static wavefield synthesis. However, investigating \cref{eq:p_ibc}, we recognize that for IBCs the wavefield to be synthesized depends on the currently propagating wavefield that will interact with the boundary to be immersed. Once the IBCs actively modify the physically propagating wavefield at any point on the boundary, this virtual scattering is then free to experience any physical wave propagation phenomena and again interact any number of times with the virtual scatterer.

\subsection{Cloning an object}
From \cref{eq:extrap} we see that a set of \gfs{} is required to replace the medium within $V^\mathrm{I}$ with a desired virtual medium. These \gfs{} are used to extrapolate particle velocity and pressure from the recording to the emitting surface ${\{p,v\}(\so) \overset{G}\mapsto \{p,v\}(\si)}$. Generally they can be of analytical origin, retrieved from numerical simulations or physically measured. In \citet{becker_broadband_2021} the \gfs{} were obtained by finite-element modeling of impulsive monopole and dipole sources on $\si$ to get the pressure and particle velocity on $\so$ and applying source-receiver reciprocity. Instead of numerical \gfs{}, the physically recorded scattering \gfs{} of an object can also be used. The outlined holography approach then allows the object's exact reproduction in different physical or even numerical environments, hence, cloning the object acoustically.

Comparing the configurations for scattering \gf{} retrieval using MDD in \cref{fig:MDDconfig} and real-time broadband holography in \cref{fig:holgraphyConfig}, we see that they are nearly identical. Both contain two mathematically closed, sound transparent surfaces. For MDD the inner surface $\si$ records the wavefield $\{p,v\}(\si)$ whereas for holography it emits the IBC source field $\{q,f\}(\si)$. Therefore, if $\si$ is instrumented with suitable transducers that can be used as both sensors and actuators, the two configurations can exactly be mapped onto each other.

In this case, cloning can be effected in two simple steps: In the first step, a physical scattering object is placed in $V^\mathrm{I}$ and illuminated by a set of broadband sources from $V^\mathrm{ext}$ while the individual wavefields are recorded on $\si$ and $\so$ (using the transducers as receivers). The incident wavefield $\{p,v\}_\mathrm{in}(\so)$ is obtained by wavefield decomposition and the scattering \gfs{} between all pairs of receivers spanning the two surfaces are retrieved using MDD.

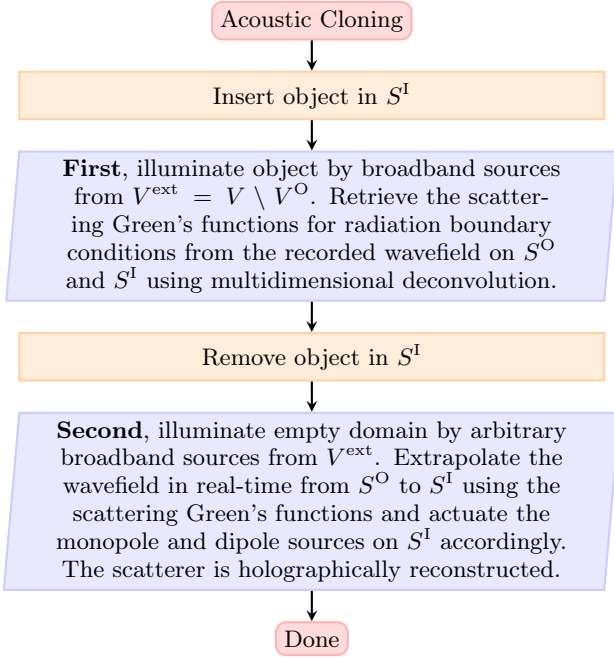
\begin{figure}
	\centering
	\begin{tikzpicture}[
	anchor=center,
	start chain=going below,
	every join/.style={arrow},
	node distance=0.4cm
	]
	\small
	
	\node [startstop,on chain,join] {Acoustic Cloning};
	
	\node [action, on chain,join] {Insert object in $\si$};
	
	\node [data, on chain,join] {\textbf{First}, illuminate object by broadband sources from $V^\mathrm{ext} = V \setminus V^\mathrm{O}$. Retrieve the scattering \gfs{} for radiation boundary conditions from the recorded wavefield on $\so$ and $\si$ using multidimensional deconvolution.};
	
	\node [action, on chain,join] {Remove object in $\si$};
	
	\node [data, on chain,join] {\textbf{Second}, illuminate empty domain by arbitrary broadband sources from $V^\mathrm{ext}$. Extrapolate the wavefield in real-time from $\so$ to $\si$ using the scattering \gfs{} and actuate the monopole and dipole sources on $\si$ accordingly. The scatterer is holographically reconstructed.};
	
	\node [startstop, on chain,join] {Done};
\end{tikzpicture} 	\caption{Visualization of the two-step process for acoustic cloning. Both steps are further subdivided visually in \cref{app:mdd}.}
	\label{fig:flowclone}
\end{figure}

In the second step, the scatterer is removed and the retrieved scattering \gfs{} (for radiation conditions) are used to reconstruct the scatterer holographically, either experimentally in real-time, or numerically. The cloned scatterer can then be subject to further physical or numerical scattering experiments. There, the clone reproduces the response of the scatterer for any incident wavefield and order of interaction exactly, as if the original object were present. The cloning process is summarized in the flowchart in \cref{fig:flowclone}.
The configurations in \cref{fig:MDDconfig} and \cref{fig:holgraphyConfig} are carefully chosen to ensure that both \gf{} retrieval using MDD and real-time holography can be realized experimentally using the same geometries. The reason for not exploring the limiting case of $\si$ approaching $\so$ (as done by \citet{liClosedapertureUnboundedAcoustics2021}) in the scattering \gf{} retrieval thus becomes apparent. The distance between $\si$ and $\so$ should exceed the minimal numerical lag necessary to record, extrapolate and emit the IBC source signals in real-time to reproduce the scatterer holographically. In addition, this approach completely avoids the (integrable) singularities that arise when collapsing $\si$ onto $\so$.

\section{Experimental setup}
\label{sec:ExpSetup}
The experimental setup used to demonstrate the proposed cloning methodology closely follows the setup used by \citet{becker_broadband_2021} for real-time broadband holography. It is based on an air-filled acoustic waveguide composed of two parallel polymethyl methacrylate (PMMA) plates separated by $2.5$~cm, as depicted in \cref{fig:setup}. Below the cutoff frequency of the fundamental mode ($\approx 6.9$~kHz for our experiments), the waveguide supports purely 2D wave propagation. The experimental domain is bounded by a 3D-printed cylinder of the synthetic resin VeroWhite with a radius of 66.2~cm. Because of its rigidity it constitutes a strongly reflective boundary.

To record the full wavefield on $\so$ with a radius $r^\mathrm{O}=36.3$~cm, we use two circular auxiliary surfaces $\so_\pm$ of pressure microphones at the radii $r^\mathrm{O}_\pm = r^\mathrm{O}\pm \Delta r$ and the finite-difference spacing $\Delta r= 1~$cm (cf. \cref{fig:setup}). Each auxiliary surface consists of $N^\mathrm{rec} = 114$ evenly spaced electret condenser microphones (COM-08635 with preamplifier BOB-12758, SparkFun electronics).  All microphones are mounted flush with the inside of the PMMA plate to minimize their impact on the pressure field. The wavefield $\{p,v\}(\mathbf{x}^o \in \so)$ can then be estimated from $p(\mathbf{x}^o_\pm \in \so_\pm)$ using \cite{beckerImmersiveWavePropagation2018}
\begin{equation}
	\begin{bmatrix*}[c]
		p \\
		\Delta p
	\end{bmatrix*}(\mathbf{x}^o)\approx
\frac{1}{2}
	\begin{bmatrix*}[r]
		1 & 1 \\
		1 & -1
	\end{bmatrix*}
	\begin{bmatrix*}[c]
	p(\mathbf{x}^o_+)\\
	p(\mathbf{x}^o_-)
\end{bmatrix*},
\label{eq:pv}
\end{equation}
and integrating the finite-difference approximation ${\Delta p \approx \Delta r \rho\, dv/dt}$ to get $v$. For real-time wavefield extrapolation in the holographic reconstruction, the pressure on the recording surface is therefore estimated by the mean pressure of the two adjacent auxiliary surfaces. The particle velocity is estimated by the time integration of the finite-difference given by the auxiliary pressure recordings.

For the inner surface $\si$, transceivers (simultaneous transmitter and receiver) would be preferred since, in that case, no reconfiguration between the scattering \gf{} retrieval and real-time holography would be necessary. Nevertheless, the results presented here are obtained by first occupying $\si$ with pressure microphones during the \gf{} retrieval step and subsequently exchanging them with loudspeakers [PUI Audio, AS01508MR-R; see \cref{fig:setupPicture}] for the real-time holography step.

To implement the inner surface $\si$ with radius $r^\mathrm{I}=8.4\mathrm{cm}$ in practice, we again use two adjacent auxiliary surfaces at $r^\mathrm{I}_\pm=r^\mathrm{I}\pm \Delta r$ each containing $N^\mathrm{emt} = 18$ evenly spaced receivers (or speakers in the holography step). In the first recording step, we can simply use \cref{eq:pv} to get $\{p,v\}(\si)$ so that we can extract the full set of \gf{} $G\equiv \{G^{p\mid q},G^{p\mid f},G^{v\mid q},G^{v\mid f}\}$ through MDD as described above. In the second step, we require monopole $q$ and dipole $f$ sources on this inner surface $\{q,f\}(\si)$ according to \cref{eq:p_ibc}. The monopole $q$ source is then given by acting on both speakers with the same source signal and the dipole $f$ by acting with opposing source signals. A detailed description of the source implementation is given in \Cref{app:setup}.

The $30$-cm distance from $\so$ to $\si$ allows a maximum total latency of about 870~$\mu$s (for experiments in air), of which 200~$\mu$s are needed for the wavefield extrapolation. The remaining 670~$\mu$s are available for real-time hardware corrections and estimation of normal particle velocity from pressure recordings. The real-time extrapolation algorithm is implemented on a massively parallelized, low-latency control system enabled by 500 National Instruments FlexRIO field-programmable gate arrays (\mbox{FPGAs}). The system supports simultaneous recording of up to 800 analog input channels and simultaneous emission of 800 analog output channels while operating at a 20-kHz sampling rate. For more details about the system architecture, we refer to \citet{beckerImmersiveWavePropagation2018}.

\section{Results}
\subsection{Retrieving \gfs{}}
\label{sec:mddresults}
\begin{figure*}
	\begin{subfigure}[c]{0.5\textwidth}
		\includegraphics[width=\textwidth,trim={6.4cm 0 5.5cm 1.8cm}, clip]{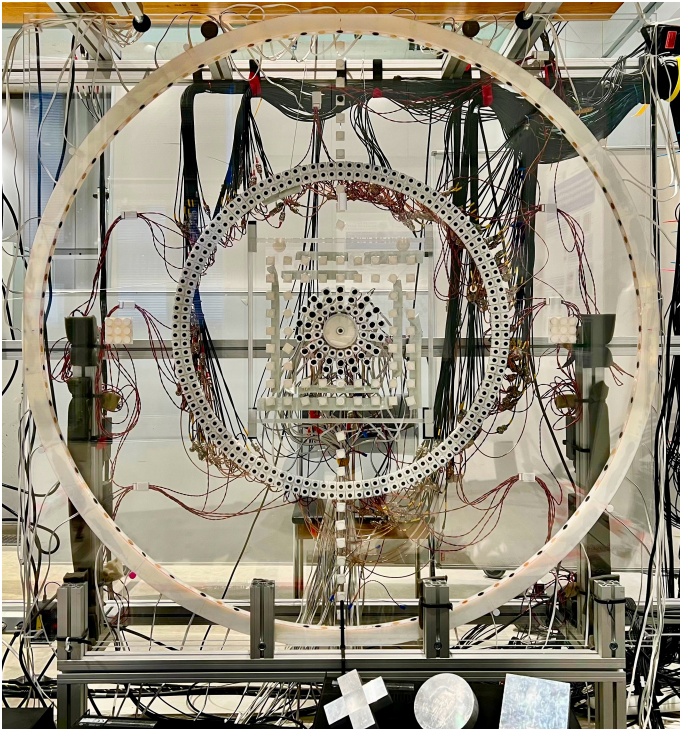}
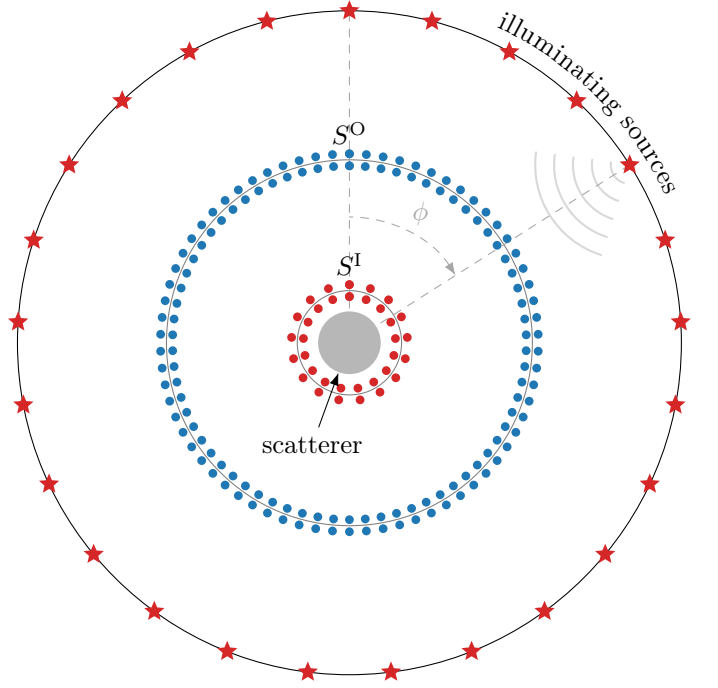
		\caption{Physical setup}
		\label{fig:setupPicture}
	\end{subfigure}
    \hfill
	\begin{subfigure}[c]{0.49\textwidth}
		\begin{tikzpicture}[
		line cap=round,
		line join=round,
		x=1cm,y=1cm,
		scale = 1
	]
	
	\pgfmathsetlengthmacro{\rdomain}{0.6621*0.8 cm}
	
\pgfmathsetmacro{\s}{\textwidth/\rdomain/2}

	\pgfmathsetlengthmacro{\Rsource}{0.007*\s cm}
	\pgfmathsetlengthmacro{\Rscat}{0.05*\s cm}

	\pgfmathsetlengthmacro{\RemtInner}{0.0737*\s cm}
	\pgfmathsetlengthmacro{\RemtOuter}{0.0927*\s cm} \pgfmathsetlengthmacro{\Remt}{(\RemtInner+\RemtOuter)/2}

	\pgfmathsetlengthmacro{\RrecInner}{0.3529*\s*0.8 cm}
	\pgfmathsetlengthmacro{\RrecOuter}{0.3769*\s*0.8 cm} \pgfmathsetlengthmacro{\Rrec}{(\RrecInner+\RrecOuter)/2}

	\pgfmathsetlengthmacro{\Rdomain}{\rdomain*\s}

	\pgfmathsetmacro{\illSourceAngle}{90-360/25*4}

	\coordinate (A) at (0,0);

	\newcommand{\tstar}[5]{\pgfmathsetmacro{\starangle}{360/#3}
		\path[#5] (#4:#1)
		\foreach \x in {1,...,#3}
			{ -- (#4+\x*\starangle-\starangle/2:#2) -- (#4+\x*\starangle:#1)
			}
		-- cycle;
	}

	\draw[gray] circle [radius=\Remt];
	\draw[gray] circle [radius=\Rrec];

	\draw[] circle [radius=\Rdomain];
	\path[decorate,decoration={text effects along path, text={illuminating sources},text align=center,text along path,raise=0.22cm}] (90:\Rdomain) arc (90:0:\Rdomain);

	\draw[gray!30,thick,decorate,decoration={expanding waves,angle=40,segment length=0.25cm}] (\illSourceAngle:\Rdomain) -- (\illSourceAngle:\Rdomain*0.7) ;

\draw[gray!70,dashed] (\illSourceAngle:\Rdomain) coordinate (A) -- (0,0) coordinate (B)
	-- (90:\Rdomain) coordinate (C)
	pic ["$\phi$", draw, Latex-,angle eccentricity=1.17,angle radius=1.07*(\Remt+\Rrec)/2,] {angle};

\fill[\colscat!80] circle [radius=\Rscat];
	\node[circle,minimum height=\Rscat*2] (scatterer) {};

\def\nRecOuter{70}
	\foreach \r in {\RrecInner,\RrecOuter}{
			\foreach \i in {0,1,...,\nRecOuter}{
					\fill[\colrec] (90-360/\nRecOuter*\i:\r) circle [radius=\Rsource]; }; };

\def\nRecInner{17}
	\foreach \r in {\RemtInner,\RemtOuter}{
			\foreach \i in {0,1,...,\nRecInner}{
					\fill[\colsrc] (90-360/\nRecInner*\i:\r) circle [radius=\Rsource]; }; };

\foreach \i in {1,2,...,25}{
\node[sourceStar,scale=0.9] at (90-360/25*\i:\Rdomain) {};
		};

	\node[above] at (90:\RrecOuter) {$\so$};
	\node[above] at (90:\RemtOuter) {$\si$};

	\node[] at (-110:\Rrec*0.45+\Remt/2) (text) {scatterer};
	\draw [-{Latex[sep=0pt]}] (text) -- (scatterer);

\end{tikzpicture} 		\caption{Sketch of setup}
		\label{fig:setupSketch}
	\end{subfigure}
	\caption{The physical setup is depicted in \subref{fig:setupPicture} and sketched in \subref{fig:setupSketch}. The outermost circle represents a rigid boundary to the 2D waveguide. It is equipped with evenly spaced loudspeakers (red stars). The waveguide contains two sound-transparent surfaces, the outer surface $\so$ and the inner surface $\si$ (gray circles). Both surfaces are surrounded by two densely spaced auxiliary circles equipped with microphones (blue dots) or loudspeakers (red dots). For the acquisition of the MDD data, they all contain microphones, and for the holography step, $\si$ is equipped with loudspeakers (configuration as shown here). In \subref{fig:setupSketch} we placed a circular scatterer inside $\si$ (heterogeneous) whereas \subref{fig:setupPicture} is empty (homogeneous). Note that, for illustration purposes, panel \subref{fig:setupSketch} is not to scale.}
	\label{fig:setup}
\end{figure*}
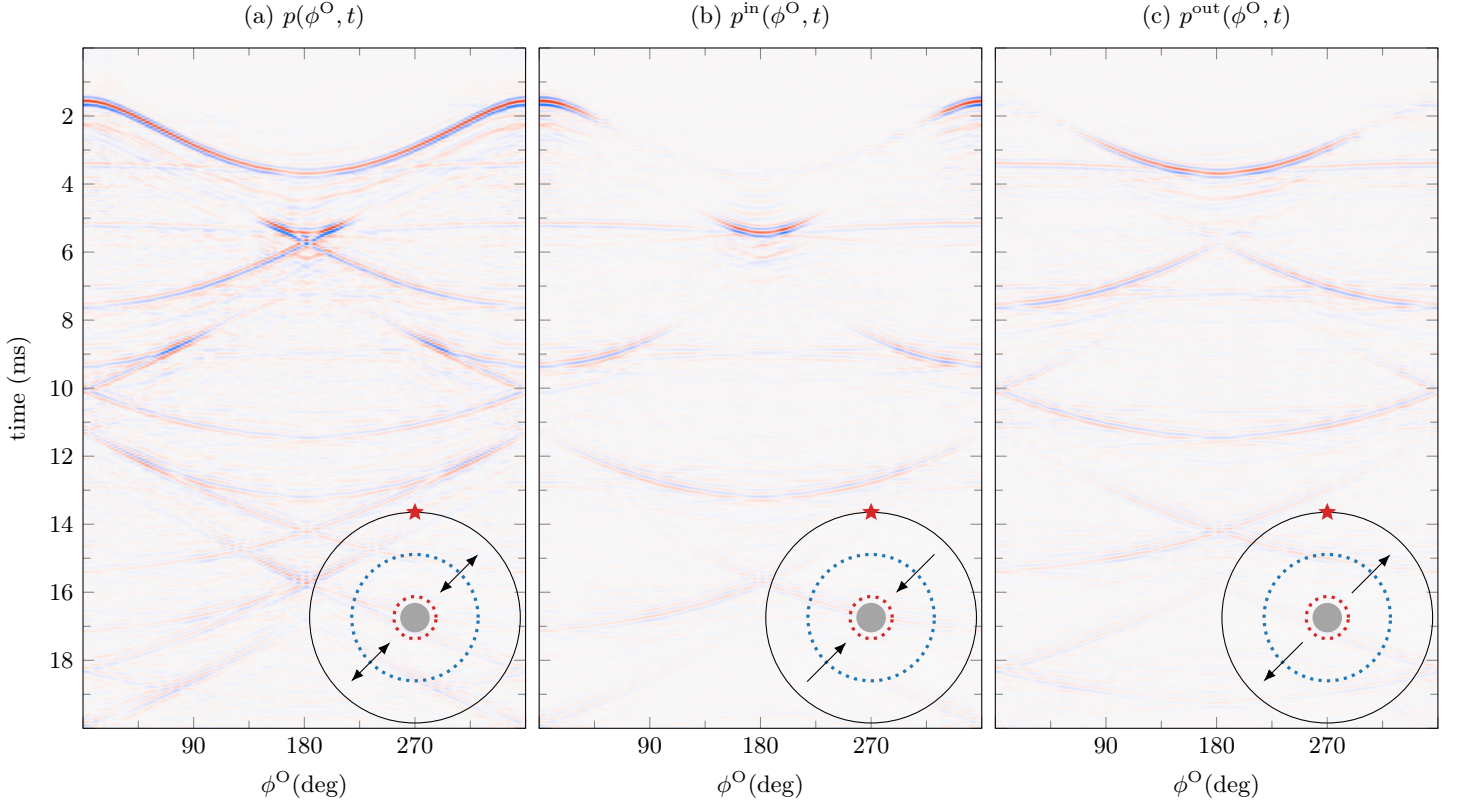
\begin{figure*}
	\newlength\figH
\newlength\figW
\newlength\yAxisW

\newlength\rInset
\setlength{\rInset}{\textwidth/4-1pt}

\setlength{\figH}{9cm}
\setlength{\figW}{(\textwidth)/3/100*98}
\setlength{\rInset}{\figW/4-2pt}

\pgfplotsset{
	sepAxis/.style = {
		width=\figW,
		height=\figH,
		at={(0\figW,0\figH)},
		xtick={0,90,180,270,360},
		xticklabels={{},90,180,270},
		scale only axis,
		minor tick num=1,
		axis on top,
		xmin=0,
		xmax=360,
		y dir=reverse,
		ymin=0,
		ymax=20,
		axis background/.style={fill=white},
		title style={font=\bfseries},
	}
}

	\begin{subfigure}[t]{\figW}
	\caption{$p(\phi^\text{O},t)$}
	
	\begin{tikzpicture}[tikzPlot]
		
		\begin{axis}[sepAxis,
			xlabel={$\phi^\text{O}(\mathrm{deg})$},
			yticklabels={{},{},2,4,...,18},
			ylabel={time (ms)},
]
			\addplot [forget plot] graphics [xmin=0, xmax=360, ymin=0, ymax=20] {6_a_sepTot.png};
		\end{axis}
	
	\tikzset{shift={(\figW-\rInset-2pt,\rInset+2pt)}}

\def\phi{45}
\def\inner{0.33}
\def\outer{0.85}
\coordinate (A) at (\phi:\rInset);

\coordinate (a1) at ($(0,0)!\inner!(A)$);
\coordinate (a2) at ($(0,0)!\outer!(A)$);

\coordinate (B) at (180+\phi:\rInset);
\coordinate (b1) at ($(0,0)!\inner!(B)$);
\coordinate (b2) at ($(0,0)!\outer9!(B)$);

\draw[black,fill=white,fill opacity=0.15] circle [radius=\rInset]; \draw[rec] circle [radius=0.6\rInset];
\draw[src] circle [radius=0.2\rInset];

\draw[draw=none,fill=\colscat] circle [radius=0.14\rInset];

\node[sourceStar,scale=0.8] at (0,\rInset) {}; 	\draw[Latex-Latex] (a1) -- (a2);
	\draw[Latex-Latex] (b1) -- (b2);
	
	\end{tikzpicture}\label{fig:sepTot}
	
\end{subfigure}
\hfill
\begin{subfigure}[t]{\figW}
	\caption{$p^\text{in}(\phi^\text{O},t)$}
	
		\begin{tikzpicture}[tikzPlot]
		
		\begin{axis}[sepAxis,
			xlabel={$\phi^\text{O}(\mathrm{deg})$},
			yticklabels={{},{},{},{},{}},
			]
			\addplot [forget plot] graphics [xmin=0, xmax=360, ymin=0, ymax=20] {6_b_sepIn.png};
		\end{axis}
		
		\tikzset{shift={(\figW-\rInset-2pt,\rInset+2pt)}}

\def\phi{45}
\def\inner{0.33}
\def\outer{0.85}
\coordinate (A) at (\phi:\rInset);

\coordinate (a1) at ($(0,0)!\inner!(A)$);
\coordinate (a2) at ($(0,0)!\outer!(A)$);

\coordinate (B) at (180+\phi:\rInset);
\coordinate (b1) at ($(0,0)!\inner!(B)$);
\coordinate (b2) at ($(0,0)!\outer9!(B)$);

\draw[black,fill=white,fill opacity=0.15] circle [radius=\rInset]; \draw[rec] circle [radius=0.6\rInset];
\draw[src] circle [radius=0.2\rInset];

\draw[draw=none,fill=\colscat] circle [radius=0.14\rInset];

\node[sourceStar,scale=0.8] at (0,\rInset) {}; 		\draw[Latex-] (a1) -- (a2);
		\draw[Latex-] (b1) -- (b2);
		
	\end{tikzpicture}\label{fig:sepIn}

\end{subfigure}
\hfill
\begin{subfigure}[t]{\figW}
	\caption{$p^\text{out}(\phi^\text{O},t)$}
	\begin{tikzpicture}[tikzPlot]
		
		\begin{axis}[sepAxis,
			xlabel={$\phi^\text{O}(\mathrm{deg})$},
			yticklabels={{},{},{},{},{}},
			]
			\addplot [forget plot] graphics [xmin=0, xmax=360, ymin=0, ymax=20] {6_c_sepOut.png};
		\end{axis}
		
		\tikzset{shift={(\figW-\rInset-2pt,\rInset+2pt)}}

\def\phi{45}
\def\inner{0.33}
\def\outer{0.85}
\coordinate (A) at (\phi:\rInset);

\coordinate (a1) at ($(0,0)!\inner!(A)$);
\coordinate (a2) at ($(0,0)!\outer!(A)$);

\coordinate (B) at (180+\phi:\rInset);
\coordinate (b1) at ($(0,0)!\inner!(B)$);
\coordinate (b2) at ($(0,0)!\outer9!(B)$);

\draw[black,fill=white,fill opacity=0.15] circle [radius=\rInset]; \draw[rec] circle [radius=0.6\rInset];
\draw[src] circle [radius=0.2\rInset];

\draw[draw=none,fill=\colscat] circle [radius=0.14\rInset];

\node[sourceStar,scale=0.8] at (0,\rInset) {}; 		\draw[-Latex] (a1) -- (a2);
		\draw[-Latex] (b1) -- (b2);
		
	\end{tikzpicture}\label{fig:sepOut}
\end{subfigure}
 	\caption{The total pressure field in \subref{fig:sepTot}, for the heterogeneous case with circular scatterer, is separated into the incident \subref{fig:sepIn} and outgoing \subref{fig:sepOut} parts. This gives the decomposition $p=p^\mathrm{in}+p^\mathrm{out}$ or $\mathrm{\subref{fig:sepTot}} = \mathrm{\subref{fig:sepIn}} + \mathrm{\subref{fig:sepOut}}$. The inset of each panel shows a schematic view of the setup, where the black arrows indicate the direction of wavefield propagation. Note that the data shown here represent only the subset for one source position of the illumination data used in multidimensional deconvolution. The source position at $\phi_\mathrm{src} = 0^\circ$ is indicated by the red star in the inset.
	}
	\label{fig:sep}
\end{figure*}
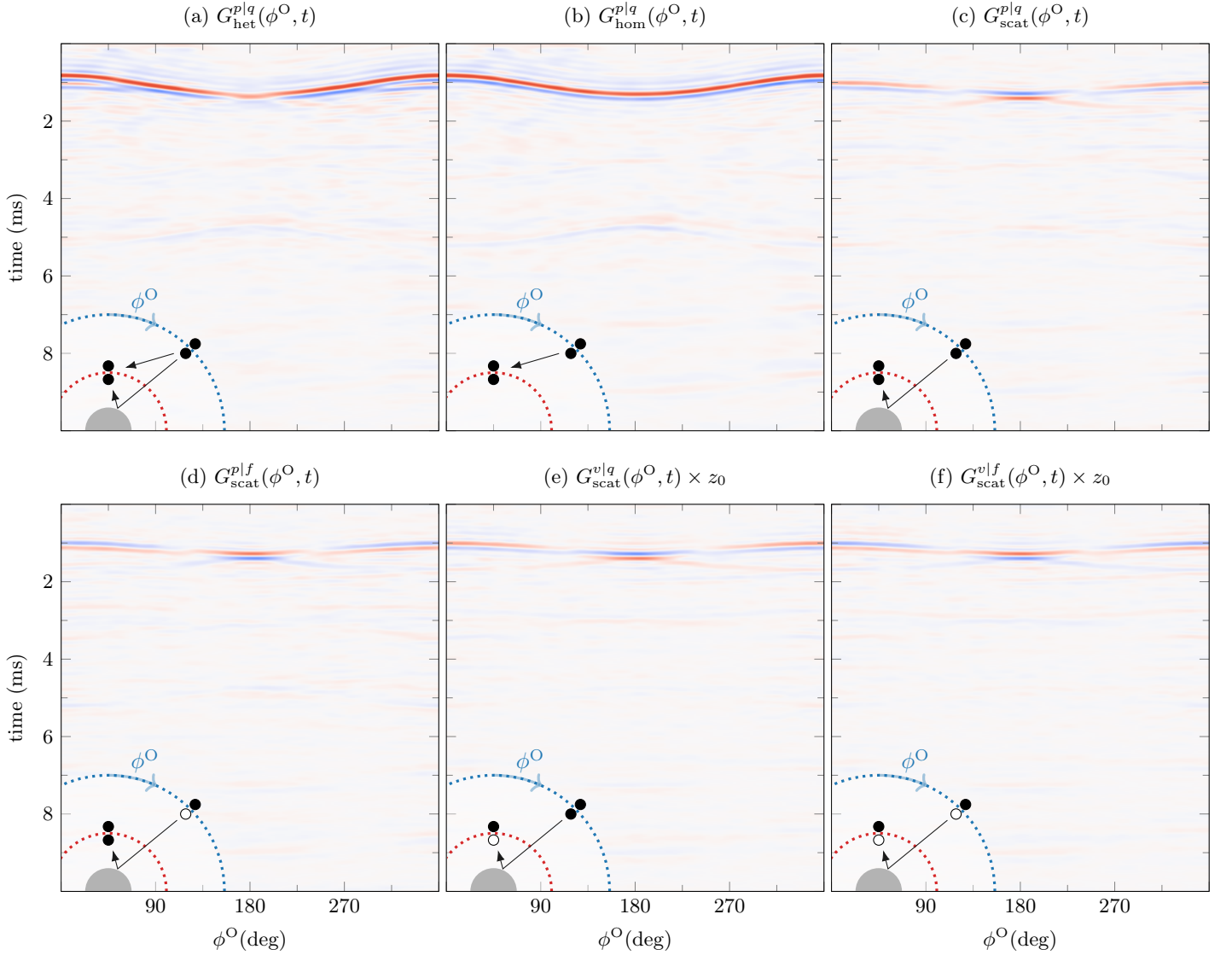
\begin{figure*}

\setlength{\rInset}{1.2cm}

\setlength{\figH}{6cm}
\setlength{\figW}{(\textwidth)/3/100*98}

\tikzset{
	rayFar/.style = {-Latex,black!90 }
}

\pgfplotsset{
	gfAxis/.style = {
		width=\figW,
		height=\figH,
		at={(0,0)},
		scale only axis,
		minor tick num=1,
		axis on top,
		xmin=0,
		xmax=360,
		xtick={0,90,180,270,360},
		y dir=reverse,
		ymin=0,
		ymax=10,
	}
}

\centering
\begin{subfigure}[c]{\figW}
\caption{$G_\text{het}^{p\mid q}(\phi^\text{O},t)$}
\begin{tikzpicture}[tikzPlot]
		\begin{axis}[gfAxis,
			xtick={0,90,180,270,360},
			xticklabels={{}},
			yticklabels={{},{},2,4,...,8},
			ylabel={time (ms)},
			title style={font=\bfseries},
]
			\addplot [forget plot] graphics [xmin=0, xmax=360, ymin=0, ymax=10] {7_a_gfHeteroPq.png};
		\end{axis}
		
\def\opacity{1}

\def\rInset{1.8cm}
\def\rSrc{0.8mm}
\def\lineWidth{0.4pt}

\def\insetShift{\figW*1/8} 

\tikzset{shift={(\insetShift,0pt)}}

\def\phiRec{45}
\def\inner{0.2}
\def\outer{0.5}
\def\sep{0.7pt}

\coordinate (pRec) at (90-\phiRec:\rInset);
\coordinate (pEmt) at (90:\outer*\rInset);
\coordinate (pRef) at (90-\phiRec/2:\inner*\rInset*1.05);

\coordinate (pRecO) at ($(pRec)+(90-\phiRec:\rSrc+\sep)$);
\coordinate (pRecI) at ($(pRec)+(270-\phiRec:\rSrc+\sep)$);

\coordinate (pEmtO) at ($(pEmt)+(90:\rSrc+\sep)$);
\coordinate (pEmtI) at ($(pEmt)+(270:\rSrc+\sep)$);

\tikzstyle{markPhi} = [postaction={decorate,decoration={markings,
		mark=at position {0.19} with {\arrow[thick]{<}},
		mark=at position {0.25} with {\arrow[thick]{|}},
}}]

\begin{scope}
	\clip (-\figW*1/8+\lineWidth,\lineWidth) rectangle (\rInset+5pt,\rInset+5pt);
	\draw[rec,fill=white,fill opacity=0.4] circle [radius=\rInset];
	\draw[src] circle [radius=\outer*\rInset] ;
	\draw[draw=none,fill=gray!60,fill opacity=\opacity] circle [radius=\inner*\rInset];
\end{scope}

\def\rAng{\rInset}
\node[rec,xshift=3pt,yshift=8pt] (angle) at (75:\rInset)  {$\phi^\text{O}$};
\draw[\colrec,very thick,opacity=0.4,->,solid] (0,\rAng) arc (90:65:\rAng) {};

\draw[fill=black] (pRecO) circle [radius=\rSrc];
\draw[fill=black] (pEmtO) circle [radius=\rSrc];

\node[circle,draw=none,minimum size=5*(\rSrc+\sep)] (recTot) at (pRec) {}; \node[circle,draw=none,minimum size=5*(\rSrc+\sep)] (emtTot) at (pEmt) {}; %

		\draw[rayFar] (recTot) -- (emtTot);
		\draw[rayFar] (recTot) -- (pRef) -- (emtTot);
		
		\draw[fill=black] (pRecI) circle [radius=\rSrc];
		\draw[fill=black] (pEmtI) circle [radius=\rSrc];
		
	\end{tikzpicture}\label{fig:gfpqhetero}
\end{subfigure}
\begin{subfigure}[c]{\figW}
	\caption{$G_\text{hom}^{p\mid q}(\phi^\text{O},t)$}
\begin{tikzpicture}[tikzPlot]
		
		\begin{axis}[gfAxis,
			xticklabels={{},{},{},{}},
			yticklabels={{},{},{},{},{}}
			title style={font=\bfseries},
			]
			\addplot [forget plot] graphics [xmin=0, xmax=360, ymin=0, ymax=10] {7_b_gfHomoPq.png};
		\end{axis}
		
\def\opacity{0}

\def\rInset{1.8cm}
\def\rSrc{0.8mm}
\def\lineWidth{0.4pt}

\def\insetShift{\figW*1/8} 

\tikzset{shift={(\insetShift,0pt)}}

\def\phiRec{45}
\def\inner{0.2}
\def\outer{0.5}
\def\sep{0.7pt}

\coordinate (pRec) at (90-\phiRec:\rInset);
\coordinate (pEmt) at (90:\outer*\rInset);
\coordinate (pRef) at (90-\phiRec/2:\inner*\rInset*1.05);

\coordinate (pRecO) at ($(pRec)+(90-\phiRec:\rSrc+\sep)$);
\coordinate (pRecI) at ($(pRec)+(270-\phiRec:\rSrc+\sep)$);

\coordinate (pEmtO) at ($(pEmt)+(90:\rSrc+\sep)$);
\coordinate (pEmtI) at ($(pEmt)+(270:\rSrc+\sep)$);

\tikzstyle{markPhi} = [postaction={decorate,decoration={markings,
		mark=at position {0.19} with {\arrow[thick]{<}},
		mark=at position {0.25} with {\arrow[thick]{|}},
}}]

\begin{scope}
	\clip (-\figW*1/8+\lineWidth,\lineWidth) rectangle (\rInset+5pt,\rInset+5pt);
	\draw[rec,fill=white,fill opacity=0.4] circle [radius=\rInset];
	\draw[src] circle [radius=\outer*\rInset] ;
	\draw[draw=none,fill=gray!60,fill opacity=\opacity] circle [radius=\inner*\rInset];
\end{scope}

\def\rAng{\rInset}
\node[rec,xshift=3pt,yshift=8pt] (angle) at (75:\rInset)  {$\phi^\text{O}$};
\draw[\colrec,very thick,opacity=0.4,->,solid] (0,\rAng) arc (90:65:\rAng) {};

\draw[fill=black] (pRecO) circle [radius=\rSrc];
\draw[fill=black] (pEmtO) circle [radius=\rSrc];

\node[circle,draw=none,minimum size=5*(\rSrc+\sep)] (recTot) at (pRec) {}; \node[circle,draw=none,minimum size=5*(\rSrc+\sep)] (emtTot) at (pEmt) {}; %
 		
		\draw[rayFar] (recTot) -- (emtTot);
		
		\draw[fill=black] (pRecI) circle [radius=\rSrc];
		\draw[fill=black] (pEmtI) circle [radius=\rSrc];
		
	\end{tikzpicture}\label{fig:gfpqhomo}
\end{subfigure}
\begin{subfigure}[c]{\figW}
	\caption{$G_\text{scat}^{p\mid q}(\phi^\text{O},t)$}
	\begin{tikzpicture}[tikzPlot]
		
		\begin{axis}[gfAxis,
			xticklabels={{},{},{},{}},
			yticklabels={{},{},{},{},{}}
			]
			\addplot [forget plot] graphics [xmin=0, xmax=360, ymin=0, ymax=10] {7_c_gfScatPq.png};
		\end{axis}
		
\def\opacity{1}

\def\rInset{1.8cm}
\def\rSrc{0.8mm}
\def\lineWidth{0.4pt}

\def\insetShift{\figW*1/8} 

\tikzset{shift={(\insetShift,0pt)}}

\def\phiRec{45}
\def\inner{0.2}
\def\outer{0.5}
\def\sep{0.7pt}

\coordinate (pRec) at (90-\phiRec:\rInset);
\coordinate (pEmt) at (90:\outer*\rInset);
\coordinate (pRef) at (90-\phiRec/2:\inner*\rInset*1.05);

\coordinate (pRecO) at ($(pRec)+(90-\phiRec:\rSrc+\sep)$);
\coordinate (pRecI) at ($(pRec)+(270-\phiRec:\rSrc+\sep)$);

\coordinate (pEmtO) at ($(pEmt)+(90:\rSrc+\sep)$);
\coordinate (pEmtI) at ($(pEmt)+(270:\rSrc+\sep)$);

\tikzstyle{markPhi} = [postaction={decorate,decoration={markings,
		mark=at position {0.19} with {\arrow[thick]{<}},
		mark=at position {0.25} with {\arrow[thick]{|}},
}}]

\begin{scope}
	\clip (-\figW*1/8+\lineWidth,\lineWidth) rectangle (\rInset+5pt,\rInset+5pt);
	\draw[rec,fill=white,fill opacity=0.4] circle [radius=\rInset];
	\draw[src] circle [radius=\outer*\rInset] ;
	\draw[draw=none,fill=gray!60,fill opacity=\opacity] circle [radius=\inner*\rInset];
\end{scope}

\def\rAng{\rInset}
\node[rec,xshift=3pt,yshift=8pt] (angle) at (75:\rInset)  {$\phi^\text{O}$};
\draw[\colrec,very thick,opacity=0.4,->,solid] (0,\rAng) arc (90:65:\rAng) {};

\draw[fill=black] (pRecO) circle [radius=\rSrc];
\draw[fill=black] (pEmtO) circle [radius=\rSrc];

\node[circle,draw=none,minimum size=5*(\rSrc+\sep)] (recTot) at (pRec) {}; \node[circle,draw=none,minimum size=5*(\rSrc+\sep)] (emtTot) at (pEmt) {}; %
 		
		\draw[rayFar] (recTot) -- (pRef) -- (emtTot);
		
		\draw[fill=black] (pRecI) circle [radius=\rSrc];
		\draw[fill=black] (pEmtI) circle [radius=\rSrc];
		
	\end{tikzpicture}\label{fig:gfpqscat}
\end{subfigure}

\begin{subfigure}[c]{\figW}
	\caption{$G_\text{scat}^{p\mid f}(\phi^\text{O},t)$}
	\begin{tikzpicture}[tikzPlot]
		
		\begin{axis}[gfAxis,
			xticklabels={{},90,180,270},
			xlabel={$\phi^\text{O}(\mathrm{deg})$},
			ylabel={time (ms)},
			yticklabels={{},{},2,4,...,8},
]
			\addplot [forget plot] graphics [xmin=0, xmax=360, ymin=0, ymax=10] {7_d_gfScatPf.png};
		\end{axis}
		
\def\opacity{1}

\def\rInset{1.8cm}
\def\rSrc{0.8mm}
\def\lineWidth{0.4pt}

\def\insetShift{\figW*1/8} 

\tikzset{shift={(\insetShift,0pt)}}

\def\phiRec{45}
\def\inner{0.2}
\def\outer{0.5}
\def\sep{0.7pt}

\coordinate (pRec) at (90-\phiRec:\rInset);
\coordinate (pEmt) at (90:\outer*\rInset);
\coordinate (pRef) at (90-\phiRec/2:\inner*\rInset*1.05);

\coordinate (pRecO) at ($(pRec)+(90-\phiRec:\rSrc+\sep)$);
\coordinate (pRecI) at ($(pRec)+(270-\phiRec:\rSrc+\sep)$);

\coordinate (pEmtO) at ($(pEmt)+(90:\rSrc+\sep)$);
\coordinate (pEmtI) at ($(pEmt)+(270:\rSrc+\sep)$);

\tikzstyle{markPhi} = [postaction={decorate,decoration={markings,
		mark=at position {0.19} with {\arrow[thick]{<}},
		mark=at position {0.25} with {\arrow[thick]{|}},
}}]

\begin{scope}
	\clip (-\figW*1/8+\lineWidth,\lineWidth) rectangle (\rInset+5pt,\rInset+5pt);
	\draw[rec,fill=white,fill opacity=0.4] circle [radius=\rInset];
	\draw[src] circle [radius=\outer*\rInset] ;
	\draw[draw=none,fill=gray!60,fill opacity=\opacity] circle [radius=\inner*\rInset];
\end{scope}

\def\rAng{\rInset}
\node[rec,xshift=3pt,yshift=8pt] (angle) at (75:\rInset)  {$\phi^\text{O}$};
\draw[\colrec,very thick,opacity=0.4,->,solid] (0,\rAng) arc (90:65:\rAng) {};

\draw[fill=black] (pRecO) circle [radius=\rSrc];
\draw[fill=black] (pEmtO) circle [radius=\rSrc];

\node[circle,draw=none,minimum size=5*(\rSrc+\sep)] (recTot) at (pRec) {}; \node[circle,draw=none,minimum size=5*(\rSrc+\sep)] (emtTot) at (pEmt) {}; %
 		
		\draw[rayFar] (recTot) -- (pRef) -- (emtTot);
		
		\draw[fill=white] (pRecI) circle [radius=\rSrc];
		\draw[fill=black] (pEmtI) circle [radius=\rSrc];
		
	\end{tikzpicture}\label{fig:gfpfscat}
\end{subfigure}
\begin{subfigure}[c]{\figW}
	\caption{$G_\text{scat}^{v\mid q}(\phi^\text{O},t)\times z_0$}
	\begin{tikzpicture}[tikzPlot]
		
		\begin{axis}[gfAxis,
			xticklabels={{},90,180,270},
			xlabel={$\phi^\text{O}(\mathrm{deg})$},
			yticklabels={{},{},{},{},{}}
			]
			\addplot [forget plot] graphics [xmin=0, xmax=360, ymin=0, ymax=10] {7_e_gfScatVq.png};
		\end{axis}
		
\def\opacity{1}

\def\rInset{1.8cm}
\def\rSrc{0.8mm}
\def\lineWidth{0.4pt}

\def\insetShift{\figW*1/8} 

\tikzset{shift={(\insetShift,0pt)}}

\def\phiRec{45}
\def\inner{0.2}
\def\outer{0.5}
\def\sep{0.7pt}

\coordinate (pRec) at (90-\phiRec:\rInset);
\coordinate (pEmt) at (90:\outer*\rInset);
\coordinate (pRef) at (90-\phiRec/2:\inner*\rInset*1.05);

\coordinate (pRecO) at ($(pRec)+(90-\phiRec:\rSrc+\sep)$);
\coordinate (pRecI) at ($(pRec)+(270-\phiRec:\rSrc+\sep)$);

\coordinate (pEmtO) at ($(pEmt)+(90:\rSrc+\sep)$);
\coordinate (pEmtI) at ($(pEmt)+(270:\rSrc+\sep)$);

\tikzstyle{markPhi} = [postaction={decorate,decoration={markings,
		mark=at position {0.19} with {\arrow[thick]{<}},
		mark=at position {0.25} with {\arrow[thick]{|}},
}}]

\begin{scope}
	\clip (-\figW*1/8+\lineWidth,\lineWidth) rectangle (\rInset+5pt,\rInset+5pt);
	\draw[rec,fill=white,fill opacity=0.4] circle [radius=\rInset];
	\draw[src] circle [radius=\outer*\rInset] ;
	\draw[draw=none,fill=gray!60,fill opacity=\opacity] circle [radius=\inner*\rInset];
\end{scope}

\def\rAng{\rInset}
\node[rec,xshift=3pt,yshift=8pt] (angle) at (75:\rInset)  {$\phi^\text{O}$};
\draw[\colrec,very thick,opacity=0.4,->,solid] (0,\rAng) arc (90:65:\rAng) {};

\draw[fill=black] (pRecO) circle [radius=\rSrc];
\draw[fill=black] (pEmtO) circle [radius=\rSrc];

\node[circle,draw=none,minimum size=5*(\rSrc+\sep)] (recTot) at (pRec) {}; \node[circle,draw=none,minimum size=5*(\rSrc+\sep)] (emtTot) at (pEmt) {}; %
 		
		\draw[rayFar] (recTot) -- (pRef) -- (emtTot);
		
		\draw[fill=black] (pRecI) circle [radius=\rSrc];
		\draw[fill=white] (pEmtI) circle [radius=\rSrc];
		
	\end{tikzpicture}\label{fig:gfvqscat}
\end{subfigure}
\begin{subfigure}[c]{\figW}
	\caption{$G_\text{scat}^{v\mid f}(\phi^\text{O},t)\times z_0$}
	\begin{tikzpicture}[tikzPlot]
		
		\begin{axis}[gfAxis,
			xticklabels={{},90,180,270},
			xlabel={$\phi^\text{O}(\mathrm{deg})$},
			yticklabels={{},{},{},{},{}}
			]
			\addplot [forget plot] graphics [xmin=0, xmax=360, ymin=0, ymax=10] {7_f_gfScatVf.png};
		\end{axis}
		
\def\opacity{1}

\def\rInset{1.8cm}
\def\rSrc{0.8mm}
\def\lineWidth{0.4pt}

\def\insetShift{\figW*1/8} 

\tikzset{shift={(\insetShift,0pt)}}

\def\phiRec{45}
\def\inner{0.2}
\def\outer{0.5}
\def\sep{0.7pt}

\coordinate (pRec) at (90-\phiRec:\rInset);
\coordinate (pEmt) at (90:\outer*\rInset);
\coordinate (pRef) at (90-\phiRec/2:\inner*\rInset*1.05);

\coordinate (pRecO) at ($(pRec)+(90-\phiRec:\rSrc+\sep)$);
\coordinate (pRecI) at ($(pRec)+(270-\phiRec:\rSrc+\sep)$);

\coordinate (pEmtO) at ($(pEmt)+(90:\rSrc+\sep)$);
\coordinate (pEmtI) at ($(pEmt)+(270:\rSrc+\sep)$);

\tikzstyle{markPhi} = [postaction={decorate,decoration={markings,
		mark=at position {0.19} with {\arrow[thick]{<}},
		mark=at position {0.25} with {\arrow[thick]{|}},
}}]

\begin{scope}
	\clip (-\figW*1/8+\lineWidth,\lineWidth) rectangle (\rInset+5pt,\rInset+5pt);
	\draw[rec,fill=white,fill opacity=0.4] circle [radius=\rInset];
	\draw[src] circle [radius=\outer*\rInset] ;
	\draw[draw=none,fill=gray!60,fill opacity=\opacity] circle [radius=\inner*\rInset];
\end{scope}

\def\rAng{\rInset}
\node[rec,xshift=3pt,yshift=8pt] (angle) at (75:\rInset)  {$\phi^\text{O}$};
\draw[\colrec,very thick,opacity=0.4,->,solid] (0,\rAng) arc (90:65:\rAng) {};

\draw[fill=black] (pRecO) circle [radius=\rSrc];
\draw[fill=black] (pEmtO) circle [radius=\rSrc];

\node[circle,draw=none,minimum size=5*(\rSrc+\sep)] (recTot) at (pRec) {}; \node[circle,draw=none,minimum size=5*(\rSrc+\sep)] (emtTot) at (pEmt) {}; %
 		
		\draw[rayFar] (recTot) -- (pRef) -- (emtTot);
		
		\draw[fill=white] (pRecI) circle [radius=\rSrc];
		\draw[fill=white] (pEmtI) circle [radius=\rSrc];
		
	\end{tikzpicture}\label{fig:gfvfscat}
\end{subfigure} 	\caption[gfs]{Green’s functions retrieved with multidimensional deconvolution from reverberative data in \cref{fig:sep}. Only the subset ${G(\phi^\mathrm{O},t\mid \phi^\mathrm{I}=0^\circ)}$,  i.e., the \gfs{} from the first (top) $\si$ position to all $\so$ positions is shown. The absence of multiple scattering underlines the effectiveness of  MDD in retrieving data for radiation boundary conditions. The top panels show $G^{p\mid q}$ for the heterogeneous case $G_\mathrm{het}$  in \subref{fig:gfpqhetero}, the homogeneous case $G_\mathrm{hom}$ in \subref{fig:gfpqhomo} and their difference $G_\mathrm{scat}$ in \subref{fig:gfpqscat} containing the scattering only, i.e., $\subref{fig:gfpqhetero}-\subref{fig:gfpqhomo}=\subref{fig:gfpqscat}$. The bottom panels show the remaining \gfs{} of this subset, namely $G^{p\mid f}$ in \subref{fig:gfpfscat}, $G^{v\mid q}$ in \subref{fig:gfvqscat} and $G^{v\mid f}$ in\subref{fig:gfvfscat}. The insets visualize the ray paths as well as source and receiver configurations of the \gfs{}. We use\monopole to denote a monopole source or pressure recording and\dipole to denote a dipole source or velocity recording. The color scale for the normal particle velocity fields is scaled by the characteristic impedance $z_0=\rho c$.} \label{fig:gfs}
\end{figure*}
A photograph of the physical experiment where the waveguide is homogeneous (empty) is shown in \cref{fig:setupPicture}. At the bottom of the picture are the three different aluminium (rigid) scatterers that we clone here. They can be inserted into the center of the waveguide in order to measure the scattering response. In \cref{fig:setupSketch} we show a sketch of the heterogeneous setup with the circular scatterer placed in the center.

For the scattering \gf{} retrieval, in the first step we illuminate the scatterer (heterogeneous) and the empty waveguide (homogeneous) from all angles. To do this, we use $N^\mathrm{src}=52$ speakers, which are equally spaced on the reflecting boundary of the waveguide at radius $r=66$~cm. We sequentially excite each speaker with a Ricker wavelet of center frequency $f_c = 3$~kHz. To improve the signal-to-noise ratio, we additionally average each acquisition over $N^\mathrm{avg} = 40$ realizations. Note that for this step the source instrument response is not deconvolved, in contrast to the subsequent holography step. Since the wavefield is recorded on $\so$ and $\si$, the MDC equations can be formulated and inverted for any sufficiently broadband excitation given sufficient reflections in time.

The wavefield separation of the pressure recordings is illustrated in \cref{fig:sep} for an illumination with the top external source $\phi_\mathrm{src}=0^\circ$. The total recorded pressure wavefield in \cref{fig:sepTot} is decomposed into the incident part, \cref{fig:sepIn}, and the outgoing part, \cref{fig:sepOut}. Note that this is the heterogeneous case and the first outgoing scattered wavefield can be seen at $t\approx3.5$~ms. This scattering is then reflected back at the boundary of the waveguide such that it reappears as incident at $t\approx5$~ms.

For all illuminating sources, the incident field together with the field recorded on $\si$ is then used to assemble the MDC equation \eqref{eq:mdc}. The MDC relationship is consequently solved as an inverse problem using an iterative damped least-squares method \cite{paigeLSQRAlgorithmSparse1982}. It carries out forward MDC in the time domain to minimize an L2 normal cost function\cite{ravasiPyLopsLinearoperatorPython2020}.

A subset of the inverted \gfs{} is shown in \cref{fig:gfs}. The top row shows the heterogeneous case, the homogeneous case, and their difference, which corresponds to the scattering \gfs{}. The heterogeneous \gfs{} contain both the direct path and the scattering path (as depicted in the inset). Since the homogeneous \gfs{} only contain the direct path, we can isolate the scattering path by subtraction. Note that the multiple boundary reflections from the wavefield-separated data in \cref{fig:sep} are effectively removed by MDD and the data for our desired unbounded state are obtained as visualized in \cref{fig:mddDesired}.

\subsection{Performing real-time holography}
\label{sec:holographyresults}
The acoustic cloning is now completed by exactly reproducing the full scattered acoustic field. We use a circular scatterer with radius $r=5$~cm, a square with side length $s=10$~cm, and a cross consisting of five unit squares of respective side length $s=3.5$~cm [see bottom of \cref{fig:setupPicture}]. To probe the holographic scattering, we again use a Ricker wavelet with center frequency $f_c = 3$~kHz emitted from the top $\phi=0^\circ$ boundary speaker. We emphasize that this is only one possibility chosen here for convenient comparison between results but does not restrict the generality of the approach, i.e., any wavefield propagating from $\so$ to $\si$ will be scattered correctly on the acoustic hologram. The bandwidth is limited only by the experimental setup and the \gfs{} obtained in it.

For inactive IBCs we retrieve the homogeneous wavefield in \cref{fig:resulthomo}, i.e., the pressure wavefield of the empty waveguide recorded on $\so$. By switching on the IBCs, we find additional scattering of the wavefield in \cref{fig:resultIbc}. Because of the geometrical spreading of the incident wavefield and the limited scattering surface area, it is difficult to assess the accuracy of the hologram from this representation. We therefore subtract the homogeneous wavefield from the heterogeneous holography [$\mathrm{\cref{fig:circIbc}}= \mathrm{\cref{fig:resultIbc}}-\mathrm{\cref{fig:resulthomo}}$] and compare it to the difference of the real scatterer [$\mathrm{\cref{fig:circScat}}= \mathrm{\cref{fig:sepTot}}-\mathrm{\cref{fig:resulthomo}}$]. The agreement is excellent for all angles $\phi^\mathrm{O}$ and all times, which implies multiple interactions between the waveguide and the cloned scatterer. Furthermore, we can compare the square scatterer in \cref{fig:squareScat,fig:squareIbc} and the cross scatterer in \cref{fig:crossScat,fig:crossIbc} to appreciate that we can clone any arbitrary complex scatterer accurately.

The first scattered wave can be seen in \cref{fig:circScat} at $t\approx3.5$~ms. This wave front then propagates to the waveguide boundary where it is reflected back and is visible again at $t\approx5$~ms. From $t\approx7$~ms onwards we start to see scattering of the boundary-reflected excitation wavefield and boundary-reflected scattering, i.e., multiple interactions. In \cref{fig:cloningresultsPolar} we sum the scattered intensities for each angle over time $I(\phi)\propto \sum_{i=1}^{N_t}p^2(\phi,t_i)$ and normalize by the maximal scattered intensity. This confirms the good agreement between the physical scatterers and their hologram.

Comparing the first scattered wave at $t\approx3.5$~ms in \cref{fig:circScat} to \cref{fig:resulthomo}, we see that the polarization of the Ricker wavelet is the same for the reflection at $\phi=0^\circ$ and is inverted for the transmission on the other side of the scatterer $\phi=180^\circ$. Since \cref{fig:circScat} is the scattered field $p_\mathrm{scat} = p_\mathrm{het} - p_\mathrm{hom}$, we can deduce that the inverted forward-scattered energy causes the attenuation in the transmitted, direct part of the wavefield. In other words, the shadow (i.e., dimming of the amplitudes) behind the scatterer is an attenuation of the direct wavefield. Because the hologram \cref{fig:circIbc} reproduces the same scattering but is physically transparent, we can deduce that our hologram effectively creates a scattered wave at $\phi=0^\circ$ and is attenuating the traversing wave at $\phi=180^\circ$.

\begin{figure*}

\pgfplotsset{
	resultsAxis/.style = {
		width=\figWResults,
		height=\figHResults,
		at={(0,0)},
		scale only axis,
		minor tick num=1,
		axis on top,
		xmin=0,
		xmax=360,
		xtick={0,90,180,270,360},
		y dir=reverse,
		ymin=0,
		ymax=12.5,
}
}

\begin{subfigure}[lc]{\figWResults}
	\caption{Hologram Off}
	\begin{tikzpicture}[tikzPlot]
		\begin{axis}[resultsAxis,
			xticklabels={{},{},{},{}},
			yticklabels={{},{},2,4,...,12},
			ylabel={time (ms)},
			]
			\addplot [forget plot] graphics [xmin=0, xmax=360, ymin=0, ymax=12.5] {8_a_circLiveHomo.png}; \end{axis}

\def\phi{45}
\def\inner{0.33}
\def\outer{0.85}
\def\phiRec{45}
\def\rOuter{0.65*\rInsetResults}
\def\rInner{0.3*\rInsetResults}
\def\rAng{\rOuter}
\def\sep{0.7pt}

\def\rscat{0.7*\rInner}
\pgfmathsetmacro{\sscat}{\rscat}

\tikzset{shift={(\figWResults/8,\rInner+2pt)}}

\coordinate (pRec) at (90-\phiRec:\rOuter);
\coordinate (pRef) at (90-\phiRec/2:\rscat*1.18);

\def\dSquare{\rscat*1.2}
\def\sSquare{\dSquare/1.4}
\def\angRot{60}

\pgfmathsetmacro{\crossScale}{\sscat/1cm/3*1.12}

\begin{scope}
	\clip (-\figWResults/8,-\rInner-2pt) rectangle (\rInsetResults+2pt,\rInsetResults+5pt);
	\draw[black,fill=white,fill opacity=0.4] circle [radius=\rInsetResults];
	\draw[rec] circle [radius=\rOuter];
	\draw[src] circle [radius=\rInner];
	
\end{scope}

\node[sourceStar,scale=0.8] (pSrc) at (0,\rInsetResults) {};

\tikzstyle{ibcDashed}=[densely dashed,draw = \colscatibc,very thick] 		
		\draw[ray] (pSrc) -- (pRec);
		
	\end{tikzpicture}\label{fig:resulthomo}
\end{subfigure}
\begin{subfigure}[c]{\figWResults}
	\caption{Hologram On}
	
	\begin{tikzpicture}[tikzPlot]
		
		\begin{axis}[resultsAxis,
			xticklabels={{},{},{},{}},
			yticklabels={{},{},{},{},{}},
			]
			\addplot [forget plot] graphics [xmin=0, xmax=360, ymin=0, ymax=12.5] {8_b_circLiveHetero.png}; \end{axis}

\def\phi{45}
\def\inner{0.33}
\def\outer{0.85}
\def\phiRec{45}
\def\rOuter{0.65*\rInsetResults}
\def\rInner{0.3*\rInsetResults}
\def\rAng{\rOuter}
\def\sep{0.7pt}

\def\rscat{0.7*\rInner}
\pgfmathsetmacro{\sscat}{\rscat}

\tikzset{shift={(\figWResults/8,\rInner+2pt)}}

\coordinate (pRec) at (90-\phiRec:\rOuter);
\coordinate (pRef) at (90-\phiRec/2:\rscat*1.18);

\def\dSquare{\rscat*1.2}
\def\sSquare{\dSquare/1.4}
\def\angRot{60}

\pgfmathsetmacro{\crossScale}{\sscat/1cm/3*1.12}

\begin{scope}
	\clip (-\figWResults/8,-\rInner-2pt) rectangle (\rInsetResults+2pt,\rInsetResults+5pt);
	\draw[black,fill=white,fill opacity=0.4] circle [radius=\rInsetResults];
	\draw[rec] circle [radius=\rOuter];
	\draw[src] circle [radius=\rInner];
	
\end{scope}

\node[sourceStar,scale=0.8] (pSrc) at (0,\rInsetResults) {};

\tikzstyle{ibcDashed}=[densely dashed,draw = \colscatibc,very thick] 		
		\draw[ray] (pSrc) -- (pRec);
		\draw[ray] (pSrc) -- (pRef) -- (pRec);
		\draw[ibcDashed] circle [radius=\rscat];
		
	\end{tikzpicture}\label{fig:resultIbc}
\end{subfigure}
\begin{subfigure}[c]{\figWResults}
	\caption{Circular Scatterer}
	\begin{tikzpicture}[tikzPlot]
		
		\begin{axis}[resultsAxis,
			xticklabels={{},{},{},{}},
			yticklabels={{},{},{},{},{}},
			]
			\addplot [forget plot] graphics [xmin=0, xmax=360, ymin=0, ymax=12.5] {8_c_circOfflineScat.png};
		\end{axis}

\def\phi{45}
\def\inner{0.33}
\def\outer{0.85}
\def\phiRec{45}
\def\rOuter{0.65*\rInsetResults}
\def\rInner{0.3*\rInsetResults}
\def\rAng{\rOuter}
\def\sep{0.7pt}

\def\rscat{0.7*\rInner}
\pgfmathsetmacro{\sscat}{\rscat}

\tikzset{shift={(\figWResults/8,\rInner+2pt)}}

\coordinate (pRec) at (90-\phiRec:\rOuter);
\coordinate (pRef) at (90-\phiRec/2:\rscat*1.18);

\def\dSquare{\rscat*1.2}
\def\sSquare{\dSquare/1.4}
\def\angRot{60}

\pgfmathsetmacro{\crossScale}{\sscat/1cm/3*1.12}

\begin{scope}
	\clip (-\figWResults/8,-\rInner-2pt) rectangle (\rInsetResults+2pt,\rInsetResults+5pt);
	\draw[black,fill=white,fill opacity=0.4] circle [radius=\rInsetResults];
	\draw[rec] circle [radius=\rOuter];
	\draw[src] circle [radius=\rInner];
	
\end{scope}

\node[sourceStar,scale=0.8] (pSrc) at (0,\rInsetResults) {};

\tikzstyle{ibcDashed}=[densely dashed,draw = \colscatibc,very thick] 		
		\draw[ray] (pSrc) -- (pRef) -- (pRec);
		\draw[draw=\colscat,very thick,fill=\colscat] circle [radius=\rscat];
		
	\end{tikzpicture}\label{fig:circScat}
\end{subfigure}
\begin{subfigure}[c]{\figWResults}
	\caption{Circular Hologram}
	\begin{tikzpicture}[tikzPlot]
		
		\begin{axis}[resultsAxis,
			xticklabels={{},{},{},{}},
			yticklabels={{},{},{},{},{}},
			]
			\addplot [forget plot] graphics [xmin=0, xmax=360, ymin=0, ymax=12.5] {8_d_circLiveIbc.png};
		\end{axis}

\def\phi{45}
\def\inner{0.33}
\def\outer{0.85}
\def\phiRec{45}
\def\rOuter{0.65*\rInsetResults}
\def\rInner{0.3*\rInsetResults}
\def\rAng{\rOuter}
\def\sep{0.7pt}

\def\rscat{0.7*\rInner}
\pgfmathsetmacro{\sscat}{\rscat}

\tikzset{shift={(\figWResults/8,\rInner+2pt)}}

\coordinate (pRec) at (90-\phiRec:\rOuter);
\coordinate (pRef) at (90-\phiRec/2:\rscat*1.18);

\def\dSquare{\rscat*1.2}
\def\sSquare{\dSquare/1.4}
\def\angRot{60}

\pgfmathsetmacro{\crossScale}{\sscat/1cm/3*1.12}

\begin{scope}
	\clip (-\figWResults/8,-\rInner-2pt) rectangle (\rInsetResults+2pt,\rInsetResults+5pt);
	\draw[black,fill=white,fill opacity=0.4] circle [radius=\rInsetResults];
	\draw[rec] circle [radius=\rOuter];
	\draw[src] circle [radius=\rInner];
	
\end{scope}

\node[sourceStar,scale=0.8] (pSrc) at (0,\rInsetResults) {};

\tikzstyle{ibcDashed}=[densely dashed,draw = \colscatibc,very thick] 		
		\draw[ray] (pSrc) -- (pRef) -- (pRec);
		\draw[ibcDashed] circle [radius=\rscat];
		
	\end{tikzpicture}\label{fig:circIbc}
\end{subfigure}

\begin{subfigure}[c]{\figWResults}
	\caption{Square Scatterer}
	\begin{tikzpicture}[tikzPlot]
		
		\begin{axis}[resultsAxis,
			xticklabels={{},90,180,270},
			xlabel={$\phi^O(\mathrm{deg})$},
			ylabel={time (ms)},
			yticklabels={{},{},2,4,...,12},
			]
			\addplot [forget plot] graphics [xmin=0, xmax=360, ymin=0, ymax=12.5] {8_e_squareOfflineScat.png};
		\end{axis}

\def\phi{45}
\def\inner{0.33}
\def\outer{0.85}
\def\phiRec{45}
\def\rOuter{0.65*\rInsetResults}
\def\rInner{0.3*\rInsetResults}
\def\rAng{\rOuter}
\def\sep{0.7pt}

\def\rscat{0.7*\rInner}
\pgfmathsetmacro{\sscat}{\rscat}

\tikzset{shift={(\figWResults/8,\rInner+2pt)}}

\coordinate (pRec) at (90-\phiRec:\rOuter);
\coordinate (pRef) at (90-\phiRec/2:\rscat*1.18);

\def\dSquare{\rscat*1.2}
\def\sSquare{\dSquare/1.4}
\def\angRot{60}

\pgfmathsetmacro{\crossScale}{\sscat/1cm/3*1.12}

\begin{scope}
	\clip (-\figWResults/8,-\rInner-2pt) rectangle (\rInsetResults+2pt,\rInsetResults+5pt);
	\draw[black,fill=white,fill opacity=0.4] circle [radius=\rInsetResults];
	\draw[rec] circle [radius=\rOuter];
	\draw[src] circle [radius=\rInner];
	
\end{scope}

\node[sourceStar,scale=0.8] (pSrc) at (0,\rInsetResults) {};

\tikzstyle{ibcDashed}=[densely dashed,draw = \colscatibc,very thick] 		
		\draw[ray] (pSrc) -- (\sSquare/2,\sSquare*1.15) -- (pRec);
		\draw[draw=\colscat,very thick,fill=\colscat] (-90-45:\dSquare) rectangle (45:\dSquare);
		
	\end{tikzpicture}\label{fig:squareScat}
\end{subfigure}
\begin{subfigure}[c]{\figWResults}
	\caption{Square Hologram}
	\begin{tikzpicture}[tikzPlot]
		
		\begin{axis}[resultsAxis,
			xticklabels={{},90,180,270},
			xlabel={$\phi^O(\mathrm{deg})$},
			yticklabels={{},{},{},{},{}}
			]
			\addplot [forget plot] graphics [xmin=0, xmax=360, ymin=0, ymax=12.5] {8_f_squareLiveIbc.png};
		\end{axis}

\def\phi{45}
\def\inner{0.33}
\def\outer{0.85}
\def\phiRec{45}
\def\rOuter{0.65*\rInsetResults}
\def\rInner{0.3*\rInsetResults}
\def\rAng{\rOuter}
\def\sep{0.7pt}

\def\rscat{0.7*\rInner}
\pgfmathsetmacro{\sscat}{\rscat}

\tikzset{shift={(\figWResults/8,\rInner+2pt)}}

\coordinate (pRec) at (90-\phiRec:\rOuter);
\coordinate (pRef) at (90-\phiRec/2:\rscat*1.18);

\def\dSquare{\rscat*1.2}
\def\sSquare{\dSquare/1.4}
\def\angRot{60}

\pgfmathsetmacro{\crossScale}{\sscat/1cm/3*1.12}

\begin{scope}
	\clip (-\figWResults/8,-\rInner-2pt) rectangle (\rInsetResults+2pt,\rInsetResults+5pt);
	\draw[black,fill=white,fill opacity=0.4] circle [radius=\rInsetResults];
	\draw[rec] circle [radius=\rOuter];
	\draw[src] circle [radius=\rInner];
	
\end{scope}

\node[sourceStar,scale=0.8] (pSrc) at (0,\rInsetResults) {};

\tikzstyle{ibcDashed}=[densely dashed,draw = \colscatibc,very thick] 		
		\draw[ray] (pSrc) -- (\sSquare/2,\sSquare*1.15) -- (pRec);
		\draw[ibcDashed] (-90-45:\dSquare) rectangle (45:\dSquare);
		
	\end{tikzpicture}\label{fig:squareIbc}
\end{subfigure}
\begin{subfigure}[c]{\figWResults}
	\caption{Cross Scatterer}
	\begin{tikzpicture}[tikzPlot]
		
		\begin{axis}[resultsAxis,
			xticklabels={{},90,180,270},
			xlabel={$\phi^O(\mathrm{deg})$},
			yticklabels={{},{},{},{},{}}
			]
			\addplot [forget plot] graphics [xmin=0, xmax=360, ymin=0, ymax=12.5] {8_g_crossOfflineScat.png};
		\end{axis}

\def\phi{45}
\def\inner{0.33}
\def\outer{0.85}
\def\phiRec{45}
\def\rOuter{0.65*\rInsetResults}
\def\rInner{0.3*\rInsetResults}
\def\rAng{\rOuter}
\def\sep{0.7pt}

\def\rscat{0.7*\rInner}
\pgfmathsetmacro{\sscat}{\rscat}

\tikzset{shift={(\figWResults/8,\rInner+2pt)}}

\coordinate (pRec) at (90-\phiRec:\rOuter);
\coordinate (pRef) at (90-\phiRec/2:\rscat*1.18);

\def\dSquare{\rscat*1.2}
\def\sSquare{\dSquare/1.4}
\def\angRot{60}

\pgfmathsetmacro{\crossScale}{\sscat/1cm/3*1.12}

\begin{scope}
	\clip (-\figWResults/8,-\rInner-2pt) rectangle (\rInsetResults+2pt,\rInsetResults+5pt);
	\draw[black,fill=white,fill opacity=0.4] circle [radius=\rInsetResults];
	\draw[rec] circle [radius=\rOuter];
	\draw[src] circle [radius=\rInner];
	
\end{scope}

\node[sourceStar,scale=0.8] (pSrc) at (0,\rInsetResults) {};

\tikzstyle{ibcDashed}=[densely dashed,draw = \colscatibc,very thick] 		
		\draw[ray] (pSrc) -- (0.8*\rscat,0.53*\rscat) -- (pRec);
		
		\path[draw=\colscat,very thick,fill=\colscat,xscale=\crossScale,yscale=\crossScale]
		(3,1) -- (3,-1) -- (1,-1) -- (1,-3) -- (-1,-3) -- (-1,-1) -- (-3,-1) -- (-3,1) -- (-1,1) -- (-1,3) -- (1,3) -- (1,1) -- (3,1);
		
	\end{tikzpicture}\label{fig:crossScat}
\end{subfigure}
\begin{subfigure}[c]{\figWResults}
	\caption{Cross Hologram}
	\begin{tikzpicture}[tikzPlot]
		
		\begin{axis}[resultsAxis,
			xticklabels={{},90,180,270},
			xlabel={$\phi^O(\mathrm{deg})$},
			yticklabels={{},{},{},{},{}}
			]
			\addplot [forget plot] graphics [xmin=0, xmax=360, ymin=0, ymax=12.5] {8_h_crossLiveIbc.png};
		\end{axis}

\def\phi{45}
\def\inner{0.33}
\def\outer{0.85}
\def\phiRec{45}
\def\rOuter{0.65*\rInsetResults}
\def\rInner{0.3*\rInsetResults}
\def\rAng{\rOuter}
\def\sep{0.7pt}

\def\rscat{0.7*\rInner}
\pgfmathsetmacro{\sscat}{\rscat}

\tikzset{shift={(\figWResults/8,\rInner+2pt)}}

\coordinate (pRec) at (90-\phiRec:\rOuter);
\coordinate (pRef) at (90-\phiRec/2:\rscat*1.18);

\def\dSquare{\rscat*1.2}
\def\sSquare{\dSquare/1.4}
\def\angRot{60}

\pgfmathsetmacro{\crossScale}{\sscat/1cm/3*1.12}

\begin{scope}
	\clip (-\figWResults/8,-\rInner-2pt) rectangle (\rInsetResults+2pt,\rInsetResults+5pt);
	\draw[black,fill=white,fill opacity=0.4] circle [radius=\rInsetResults];
	\draw[rec] circle [radius=\rOuter];
	\draw[src] circle [radius=\rInner];
	
\end{scope}

\node[sourceStar,scale=0.8] (pSrc) at (0,\rInsetResults) {};

\tikzstyle{ibcDashed}=[densely dashed,draw = \colscatibc,very thick] 		
		\draw[ray] (pSrc) -- (0.8*\rscat,0.53*\rscat) -- (pRec);
		
		\path[ibcDashed,xscale=\crossScale,yscale=\crossScale]
		(3,1) -- (3,-1) -- (1,-1) -- (1,-3) -- (-1,-3) -- (-1,-1) -- (-3,-1) -- (-3,1) -- (-1,1) -- (-1,3) -- (1,3) -- (1,1) -- (3,1);
		
	\end{tikzpicture}\label{fig:crossIbc}
\end{subfigure} 	\caption{Recorded pressure $p(\phi^O,t)$ field for real-time cloning and reference. In \subref{fig:resulthomo} and \subref{fig:resultIbc} we show the effect of the hologram being off and on respectively. Note that in both of those cases the waveguide is homogeneous (empty). Any difference is the result of acting with the monopole and dipole sources on $\si$, i.e., the hologram. For improved visibility, we isolate the real scattering $\subref{fig:circScat}=\mathrm{\cref{fig:sepTot}} -\subref{fig:resulthomo}$ and the holographic scattering $\subref{fig:circIbc}=\subref{fig:resultIbc}-\subref{fig:resulthomo}$. Comparing the real scattering \subref{fig:circScat} with the holographic scattering \subref{fig:circIbc} confirms the validity of our approach. The same comparison is provided for the square scatterer in \subref{fig:squareScat},\subref{fig:squareIbc} and for the cross scatterer in \subref{fig:crossScat}, \subref{fig:crossIbc}.}
	\label{fig:cloningresults}
\end{figure*}
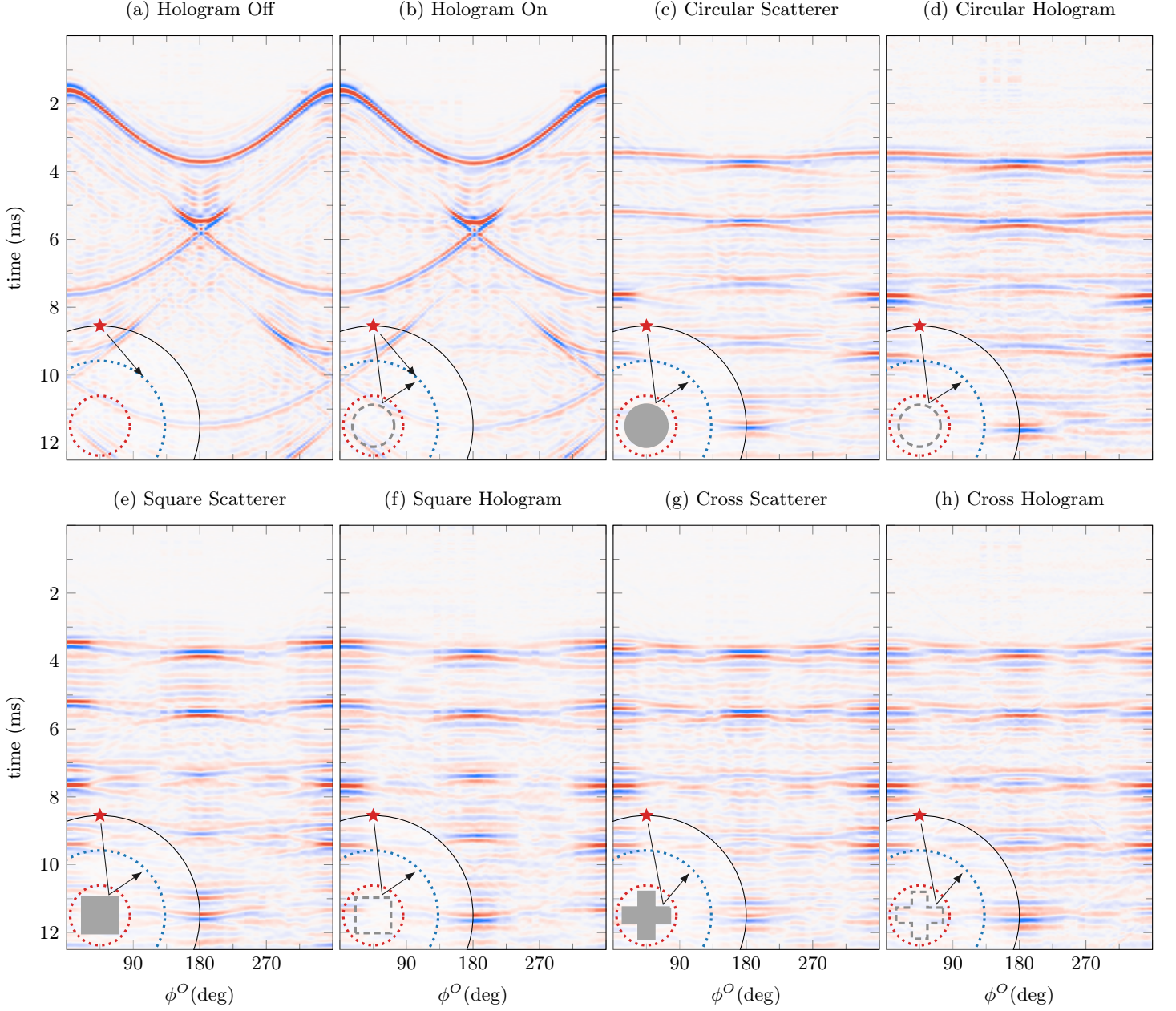
\begin{figure*}

\pgfplotstableread{data/circPolar.txt} \tableCirc
\pgfplotstableread{data/squarePolar.txt} \tableSquare
\pgfplotstableread{data/crossPolar.txt} \tableCross

\setlength{\figW}{(\textwidth)/3/100*97}

\pgfplotsset{
	gfAxis/.style = {
		width=\figW*1.08,
		rotate=-90,
		enlargelimits=false,
		x dir=reverse,
xtick={0,30,...,360},
		ytick={0,0.2,...,1},
xticklabel style={anchor=-\tick-90,font=\footnotesize,overlay},
		yticklabel style={anchor=west,font=\footnotesize,overlay},
		legend style={at={(-2.2cm,-1.8cm)},anchor=east,font=\small,overlay}, legend cell align={left},
	} }

\begin{subfigure}[t]{\figW}
	\begin{tikzpicture}[>=latex]
	 \begin{polaraxis}[gfAxis]
		\addplot+[name path = Var1,mark=none,very thick,\cI] table [x expr = {\thisrow{phi}/pi*180}, y = rScat] {\tableCirc};
		\addplot+[name path = Var1,mark=none,very thick,\cII] table [x expr = {\thisrow{phi}/pi*180}, y = rIbc] {\tableCirc};
	\end{polaraxis}
\end{tikzpicture}
	\hspace{\textheight}
	\caption{Circular}
	\label{fig:angularCirc}
\end{subfigure}
\hfill
\begin{subfigure}[t]{\figW}
	\begin{tikzpicture}[>=latex]
	\begin{polaraxis}[gfAxis]
		\addplot+[name path = Var1,mark=none,very thick,\cI] table [x expr = {\thisrow{phi}/pi*180}, y = rScat] {\tableSquare};
		\addplot+[name path = Var1,mark=none,very thick,\cII] table [x expr = {\thisrow{phi}/pi*180}, y = rIbc] {\tableSquare};
		\legend{Scatterer,Hologram};
	\end{polaraxis}
\end{tikzpicture}
	\hspace{\textheight}
	\caption{Square}
	\label{fig:angularSquare}
\end{subfigure}
\hfill
\begin{subfigure}[t]{\figW}
	\begin{tikzpicture}[>=latex]
	\begin{polaraxis}[gfAxis]
		\addplot+[name path = Var1,mark=none,very thick,\cI] table [x expr = {\thisrow{phi}/pi*180}, y = rScat] {\tableCross};
		\addplot+[name path = Var1,mark=none,very thick,\cII] table [x expr = {\thisrow{phi}/pi*180}, y = rIbc] {\tableCross};
	\end{polaraxis}
\end{tikzpicture}
	\hspace{\textheight}
	\caption{Cross}
	\label{fig:angularCross}
\end{subfigure}
\hfill 	
	\caption{Comparison of the scattered angular intensities between the real scatterers and the holographically reproduced scatterers (IBC). The sum of the scattered intensities over time from the results in \cref{fig:cloningresults} are shown. We normalize to the highest angular scattered intensity.}

	\label{fig:cloningresultsPolar}
\end{figure*}
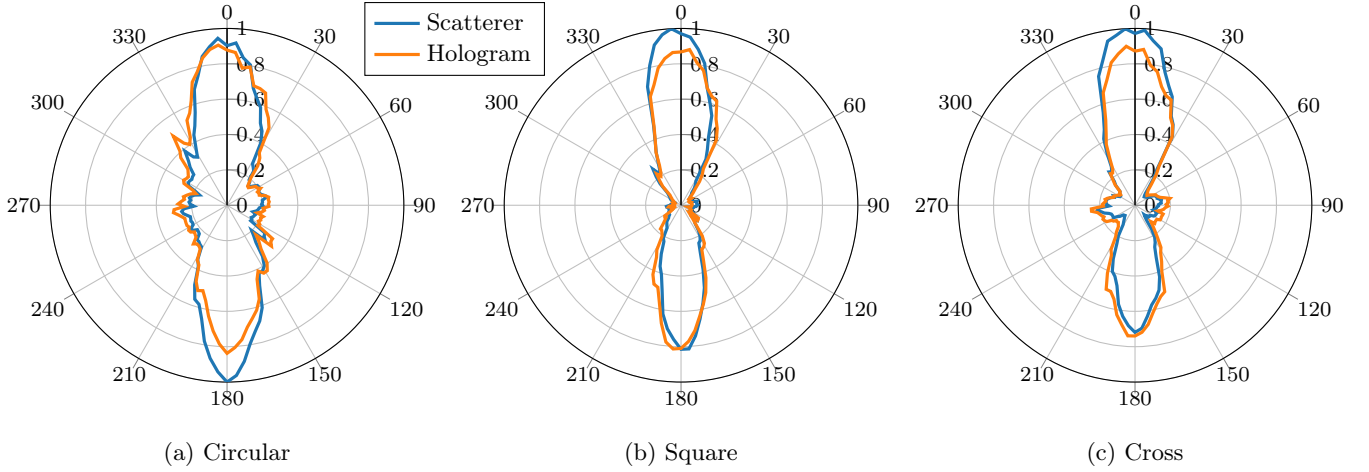

\begin{figure*}
	{\captionsetup{justification=centering}

\pgfplotsset{
	resultsAxis/.style = {
		width=\figWResults,
		height=\figHResults,
		at={(0,0)},
		scale only axis,
		minor tick num=1,
		axis on top,
		xmin=0,
		xmax=360,
		xtick={0,90,180,270,360},
		y dir=reverse,
		ymin=0,
		ymax=12.5,
}
}

\begin{subfigure}[c]{\figWResults}
	\caption{Circular Scatterer \newline \phantom{+Nothing}}
	
	\begin{tikzpicture}[tikzPlot,
		text arrow/.style={above right, align=left, yshift=4pt,xshift=-8pt, text width=1.9cm,font=\footnotesize,opacity=0.6}]
		\begin{axis}[resultsAxis,
			xticklabels={{},90,180,270},
			xlabel={$\phi^O(\mathrm{deg})$},
			yticklabels={{},{},2,4,...,12},
			ylabel={time (ms)},
			]
			\addplot [forget plot] graphics [xmin=0, xmax=360, ymin=0, ymax=12.5] {8_c_circOfflineScat.png};
			
\draw[->,opacity=0.6,black] (axis cs:30,3-0.5) -- node [ text arrow] {reflection} (axis cs:30,3);
			\draw[->,opacity=0.6,black] (axis cs:180,3.3-0.5) -- node [ text arrow] {attenuation} (axis cs:180,3.3);
		\end{axis}

\def\phi{45}
\def\inner{0.33}
\def\outer{0.85}
\def\phiRec{45}
\def\rOuter{0.65*\rInsetResults}
\def\rInner{0.3*\rInsetResults}
\def\rAng{\rOuter}
\def\sep{0.7pt}

\def\rscat{0.7*\rInner}
\pgfmathsetmacro{\sscat}{\rscat}

\tikzset{shift={(\figWResults/8,\rInner+2pt)}}

\coordinate (pRec) at (90-\phiRec:\rOuter);
\coordinate (pRef) at (90-\phiRec/2:\rscat*1.18);

\def\dSquare{\rscat*1.2}
\def\sSquare{\dSquare/1.4}
\def\angRot{60}

\pgfmathsetmacro{\crossScale}{\sscat/1cm/3*1.12}

\begin{scope}
	\clip (-\figWResults/8,-\rInner-2pt) rectangle (\rInsetResults+2pt,\rInsetResults+5pt);
	\draw[black,fill=white,fill opacity=0.4] circle [radius=\rInsetResults];
	\draw[rec] circle [radius=\rOuter];
	\draw[src] circle [radius=\rInner];
	
\end{scope}

\node[sourceStar,scale=0.8] (pSrc) at (0,\rInsetResults) {};

\tikzstyle{ibcDashed}=[densely dashed,draw = \colscatibc,very thick] 		
		\draw[ray] (pSrc) -- (pRef) -- (pRec);
		\draw[draw=\colscat,very thick, fill=\colscat] circle [radius=\rscat];
		
	\end{tikzpicture}\label{fig:circScat2}
\end{subfigure}
\begin{subfigure}[c]{\figWResults}
	\caption{Circular Hologram \newline + Gain}
	
	\begin{tikzpicture}[tikzPlot,
		text arrow/.style={above right, align=left, yshift=4pt,xshift=-8pt, text width=1.9cm,font=\footnotesize,opacity=0.6}]
		
		\begin{axis}[resultsAxis,
			xticklabels={{},90,180,270},
			xlabel={$\phi^O(\mathrm{deg})$},
			yticklabels={{},{},{},{},{}},
			]
			\addplot [forget plot] graphics [xmin=0, xmax=360, ymin=0, ymax=12.5] {10_b_circMddGain.png};
			
\draw[->,opacity=0.6,black] (axis cs:30,3-0.5) -- node [ text arrow] {amplified reflection} (axis cs:30,3);
             \draw[->,opacity=0.6,black] (axis cs:180,3.3-0.5) -- node [ text arrow] {transparent} (axis cs:180,3.3);
		\end{axis}

\def\phi{45}
\def\inner{0.33}
\def\outer{0.85}
\def\phiRec{45}
\def\rOuter{0.65*\rInsetResults}
\def\rInner{0.3*\rInsetResults}
\def\rAng{\rOuter}
\def\sep{0.7pt}

\def\rscat{0.7*\rInner}
\pgfmathsetmacro{\sscat}{\rscat}

\tikzset{shift={(\figWResults/8,\rInner+2pt)}}

\coordinate (pRec) at (90-\phiRec:\rOuter);
\coordinate (pRef) at (90-\phiRec/2:\rscat*1.18);

\def\dSquare{\rscat*1.2}
\def\sSquare{\dSquare/1.4}
\def\angRot{60}

\pgfmathsetmacro{\crossScale}{\sscat/1cm/3*1.12}

\begin{scope}
	\clip (-\figWResults/8,-\rInner-2pt) rectangle (\rInsetResults+2pt,\rInsetResults+5pt);
	\draw[black,fill=white,fill opacity=0.4] circle [radius=\rInsetResults];
	\draw[rec] circle [radius=\rOuter];
	\draw[src] circle [radius=\rInner];
	
\end{scope}

\node[sourceStar,scale=0.8] (pSrc) at (0,\rInsetResults) {};

\tikzstyle{ibcDashed}=[densely dashed,draw = \colscatibc,very thick] 		
		\draw[ray] (pSrc) -- (pRef) -- (pRec);
		\draw[ibcDashed, top color=gray, opacity=0.5] circle [radius=\rscat];
		
	\end{tikzpicture}\label{fig:resultIbcGain}
\end{subfigure}
\begin{subfigure}[c]{\figWResults}
	\caption{{Square Scatterer \newline + physical Rotation}}
	\begin{tikzpicture}[tikzPlot]
		
		\begin{axis}[resultsAxis,
			xticklabels={{},90,180,270},
			xlabel={$\phi^O(\mathrm{deg})$},
			yticklabels={{},{},{},{},{}},
			]
			\addplot [forget plot] graphics [xmin=0, xmax=360, ymin=0, ymax=12.5] {10_c_squareOfflineScatRot.png};
		\end{axis}

\def\phi{45}
\def\inner{0.33}
\def\outer{0.85}
\def\phiRec{45}
\def\rOuter{0.65*\rInsetResults}
\def\rInner{0.3*\rInsetResults}
\def\rAng{\rOuter}
\def\sep{0.7pt}

\def\rscat{0.7*\rInner}
\pgfmathsetmacro{\sscat}{\rscat}

\tikzset{shift={(\figWResults/8,\rInner+2pt)}}

\coordinate (pRec) at (90-\phiRec:\rOuter);
\coordinate (pRef) at (90-\phiRec/2:\rscat*1.18);

\def\dSquare{\rscat*1.2}
\def\sSquare{\dSquare/1.4}
\def\angRot{60}

\pgfmathsetmacro{\crossScale}{\sscat/1cm/3*1.12}

\begin{scope}
	\clip (-\figWResults/8,-\rInner-2pt) rectangle (\rInsetResults+2pt,\rInsetResults+5pt);
	\draw[black,fill=white,fill opacity=0.4] circle [radius=\rInsetResults];
	\draw[rec] circle [radius=\rOuter];
	\draw[src] circle [radius=\rInner];
	
\end{scope}

\node[sourceStar,scale=0.8] (pSrc) at (0,\rInsetResults) {};

\tikzstyle{ibcDashed}=[densely dashed,draw = \colscatibc,very thick]         
        \draw[ray] (pSrc) -- (90-\phiRec*0.6:\dSquare*0.85) -- (pRec);
		\draw[draw=\colscat,very thick,fill=\colscat, rotate=\angRot] (-90-45:\dSquare) rectangle (45:\dSquare);
        \draw[->,solid, thick,\colscat] (30:\rInner*1.5) arc (30:-30:\rInner*1.5) {};

	\end{tikzpicture}\label{fig:squareScatRot}
\end{subfigure}
\begin{subfigure}[c]{\figWResults}
	\caption{Square Hologram \newline + numerical Rotation}
	\begin{tikzpicture}[tikzPlot]
		
		\begin{axis}[resultsAxis,
			xticklabels={{},90,180,270},
			xlabel={$\phi^O(\mathrm{deg})$},
			yticklabels={{},{},{},{},{}},
			]
			\addplot [forget plot] graphics [xmin=0, xmax=360, ymin=0, ymax=12.5] {10_d_squareMddRot.png};
		\end{axis}

\def\phi{45}
\def\inner{0.33}
\def\outer{0.85}
\def\phiRec{45}
\def\rOuter{0.65*\rInsetResults}
\def\rInner{0.3*\rInsetResults}
\def\rAng{\rOuter}
\def\sep{0.7pt}

\def\rscat{0.7*\rInner}
\pgfmathsetmacro{\sscat}{\rscat}

\tikzset{shift={(\figWResults/8,\rInner+2pt)}}

\coordinate (pRec) at (90-\phiRec:\rOuter);
\coordinate (pRef) at (90-\phiRec/2:\rscat*1.18);

\def\dSquare{\rscat*1.2}
\def\sSquare{\dSquare/1.4}
\def\angRot{60}

\pgfmathsetmacro{\crossScale}{\sscat/1cm/3*1.12}

\begin{scope}
	\clip (-\figWResults/8,-\rInner-2pt) rectangle (\rInsetResults+2pt,\rInsetResults+5pt);
	\draw[black,fill=white,fill opacity=0.4] circle [radius=\rInsetResults];
	\draw[rec] circle [radius=\rOuter];
	\draw[src] circle [radius=\rInner];
	
\end{scope}

\node[sourceStar,scale=0.8] (pSrc) at (0,\rInsetResults) {};

\tikzstyle{ibcDashed}=[densely dashed,draw = \colscatibc,very thick] 		
		\draw[ray] (pSrc) -- (90-\phiRec*0.6:\dSquare*0.85) -- (pRec);
		\draw[ibcDashed, rotate=\angRot] (-90-45:\dSquare) rectangle (45:\dSquare);
        \draw[->,dashed, thick,\colscatibc] (30:\rInner*1.5) arc (30:-30:\rInner*1.5) {};
		
	\end{tikzpicture}\label{fig:squareIbcRot}
\end{subfigure}
 }
	\caption{Scattered pressure $p_\mathrm{scat}(\phi^O,t) = p_\mathrm{het}(\phi^O,t) - p_\mathrm{hom}(\phi^O,t)$ for two acoustic cyborgs. The \gfs{} for the circular and square scatterer from \cref{fig:cloningresults} are modified in the numerical domain to exhibit gain and rotation in the clone's scattering response. The circular scatterer in \subref{fig:circScat2} is cloned and the \gfs{} numerically modified \subref{fig:resultIbcGain} to include a gain in the scattering at $\phi=0^\circ$ (reflection) such that $p^\mathrm{holo}_\mathrm{scat}(0^\circ,3.3\mathrm{ms})=1.5\times p^\mathrm{real}_\mathrm{scat}(0^\circ,3.3\mathrm{ms})$, and no loss in the scattering at $\phi=180^\circ$ (transmission) such that $p^\mathrm{holo}_\mathrm{scat}(180^\circ,3.5\mathrm{ms})=0$ or equivalently $ p^\mathrm{holo}_\mathrm{het}(180^\circ,3.5\mathrm{ms})=p_\mathrm{hom}(180^\circ,3.5\mathrm{ms})$. Formally, we multiply $G_\text{scat}(\phi^\text{I})$ with the gain function $1.5(1 + \cos \phi^\text{I})/2$. The square scatterer in \subref{fig:squareScatRot} is physically rotated by $\phi^\mathrm{rot}=60^\circ$ as a reference for the same rotation performed in the numerical domain, shown in \subref{fig:squareIbcRot}.}
	\label{fig:resultsModified}
\end{figure*}
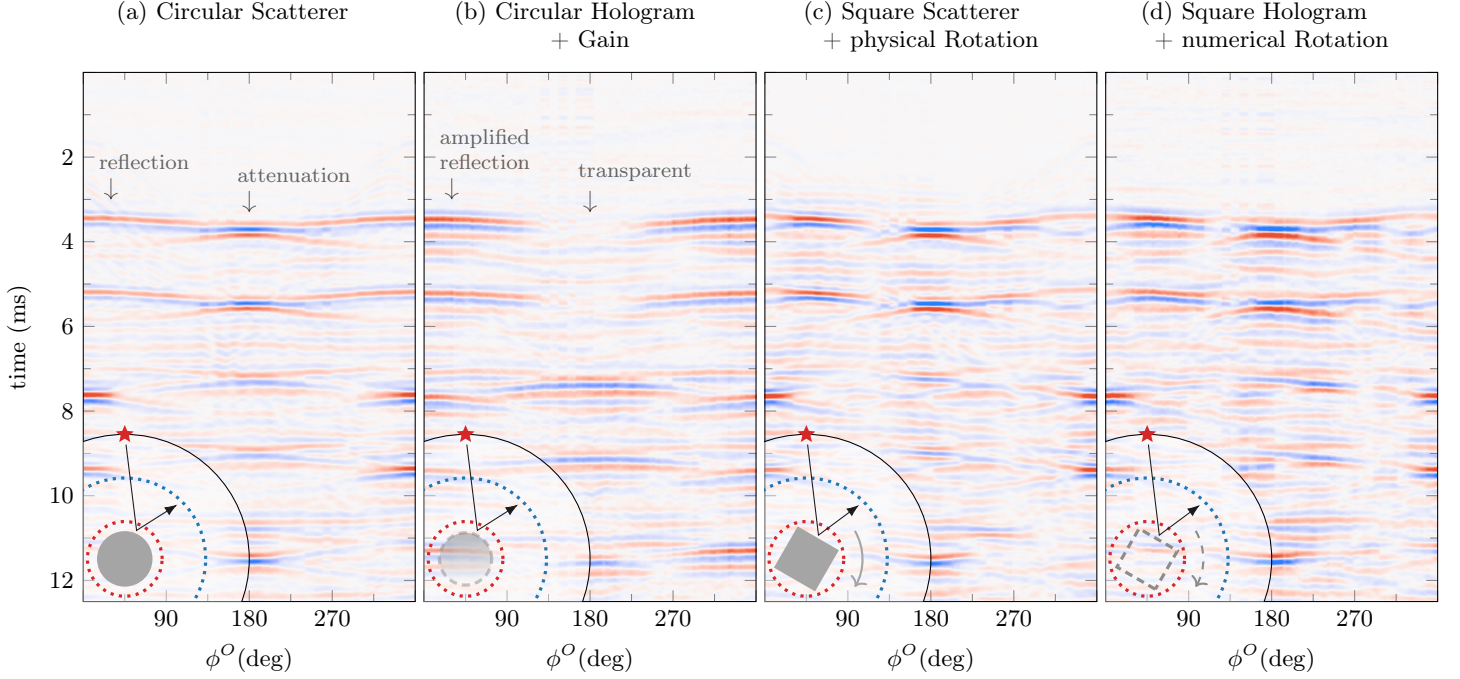

\section{Discussion}
The requirements for a broadband acoustic cloning methodology can be summarized as follows. First, the method should be general enough such that it works for arbitrary unknown scattering objects, with "unknown" meaning that no previous characterization of the scattering object is necessary in order to clone it. This implies, for example, that its dimensions, material properties, etc., do not need to be known in order to reproduce it. Second, the method should be general enough to work for a variety of experimental setups, in particular, for relatively small experimental setups that usually suffer from boundary reflections.

We further distinguish between requirements regarding the measurement of the scattering \gf{} and requirements regarding holography, i.e., real-time reproduction. First, monopole and dipole \gfs{} for radiation boundary conditions between all pairs of points on a mathematically closed recording and a mathematically closed emitting surface surrounding the object of interest are needed. Second, a spatially and temporally broadband holographic reconstruction method should enable the real-time experimental or numerical reconstruction of the full-waveform scattering of the object.

Starting with the first requirement, existing methods to acquire the impulse responses suffer from at least two significant practical problems: (1) Reflections from the experimental domain boundaries do typically exist and mask multiple scattering inside the object of interest. (2) Source and receiver instrument responses need to be measured and corrected (i.e., calibrated) individually. Conventional solutions to these problems include enlarging the experimental domain and/or padding its boundaries with materials that attenuate reflections passively or actively. Furthermore, while it is common to acquire the data using broadband sensors that have flat receiving characteristics, individual sources typically do need to emit frequency-swept signals such that the impulse response can be estimated and removed by correlation or deconvolution methods robustly. However, clearly, such solutions impose significant requirements on the experimental domain (size), and a laborious calibration process.

The methodology presented here eliminates most of these requirements, enabling acoustic cloning in relatively small and reverberant experimental setups, for which boundary reflections are usually of critical concern, with low-cost sources and (scalar) sensors requiring little calibration. These advances are in large part due to the multidimensional deconvolution employed in the scattering \gf{} retrieval and focus the attention on the remaining requirements; in particular, the low-latency acquisition, compute, and control system for the real-time holographic reconstruction. While real-time reconstruction with a high number of receiving and emitting channels poses its own set of challenges \cite{beckerImmersiveWavePropagation2018}, we note that the holographic reconstruction can also be effected in a numerical domain.

In this implementation, the terminology "immersive boundary conditions" might be somewhat misleading. Given that the boundary on which they are implemented is actually transparent, so that the acoustic impedance does not change, it is strictly speaking not a boundary. However, IBCs were derived for both Neumann ($v_n |_\mathrm{B}=0$) and Dirichlet ($p |_\mathrm{B}=0$) boundary conditions. Here we essentially use the superposition of both as is represented by the two integrands in \cref{eq:p_ibc} \cite{vanmanenBroadbandCloakingHolography2015}.

An immediate consequence of cloning is that real-time broadband cloaking appears as a byproduct when using the homogeneous \gfs{} in the holography step \cite{becker_broadband_2021,vanmanenBroadbandCloakingHolography2015}. They contain only the direct waves propagating from the outer to the inner surface $\si$. On $\si$ the ingoing field can simply be canceled and the outgoing field recreated on the other side. No energy can reach an object within $\si$ and therefore no scattering is possible inside $\si$ . At the same time, the homogeneous wave propagation is fully recovered by the numerical propagation through $\si$. Furthermore, any real scatterer can be transformed into any other scatterer by using the heterogeneous \gfs{}. Since they contain the direct and scattered waves, they allow one to simultaneously cloak the physically present scatterer and represent any other scatterer.

We fully recover the scattering object experimentally and are able to modify it to behave in a nonphysical way due to the active sources. For example, gain components are simply a numerical multiplication, phase shifts an interpolation, and any (nonphysical) combination of them can be examined to interact with physically propagating waves.

The cloning methodology is not restricted to recovering the physical scattering of an absent object only, but it also allows for enhancements or modifications of the clone. First, the clone can be modified through physical feasible operations like translations or rotations to observe a different configuration of the initial scatterer. Second, it can also include physical infeasible operations like a gain component or arbitrary propagation velocities. Because of the implementation through active sources, unphysical scattering of the object can be recovered experimentally. Two exemplary experimental scattering results for a virtual gain component and a virtually rotated scatterer are presented in \cref{fig:resultsModified}.

Compare the original circular scattering \cref{fig:circScat2} to the modified clone's scattering which includes a directional gain component \cref{fig:resultIbcGain}. The scattered amplitude is modified to exhibit the angle dependency $1.5(1 + \cos \phi^\text{I})/2$. We can interpret this modified scatterer as scattering a wave with an amplified reflection $R_\mathrm{holo}(0^\circ)=1.5 R_\mathrm{circ}(0^\circ)$ in the backwards direction while being transparent to the passing wave in the forward direction, $T_\mathrm{holo}(180^\circ)=1$. In other words, the scattered energy is not deducted from the transmitted wavefield that caused the scattering.
In \cref{fig:squareScatRot} we show the scattering for the square which is physically rotated by $\phi^\mathrm{rot}=60^\circ$. The same rotation is performed in the virtual domain so that the hologram creates the same rotated scattering in \cref{fig:squareIbcRot}. For this case we can simply shift the mapping of the \gfs{} by $N/360\times60$, i.e. three sources and 19 receivers. However, any noninteger shift can be achieved by interpolation just like a mirroring operation or any other conceivable reshuffling of the spatial reflections.

The generality of the approach inspires numerous future applications, starting from creating numerical twins of real-life objects without the need for an anechoic chamber. Such a numerical representation can then be examined, modified, and copied as many times as needed. For example, the \gfs{} of the unit cell of an acoustic metamaterial could be retrieved and modified to include spatially dependent gain. Furthermore, one can immediately examine a finite or even infinite repetition number of the unit cell (through periodic boundary conditions) without the imperfections which arise during physical reproduction of the unit cell.

\section{Conclusion}
We have successfully shown how to clone an acoustic scatterer through a simple two-step process. First, the acoustic scattering properties of an arbitrary object were fully retrieved from recorded illumination data only. These illumination data simply need to cover the desired temporal bandwidth and scatterer angles but are not limited by boundary reverberations or illumination source characteristics. We recovered a set of \gfs{} describing the object's scattering response. It was then shown how to use this set of \gfs{} to holographically recreate the acoustic scattering of the object in its absence. This was implemented in real-time using immersive boundary conditions and the results were compared to the real scatterer. An excellent agreement has been found, confirming the accuracy and generality of the proposed approach.

\begin{acknowledgments}
We would like to thank Thomas Haag and Christoph Bärlocher for their help on the hardware and software side of the experimental setup. Furthermore, we appreciated the discussions with Henrik R. Thomsen during this work. This project has received funding from SNF grant 197182. More  information about our work can be found at \url{eeg.ethz.ch/research/centre-immersive}.
\end{acknowledgments}

\appendix

\section{Wavefield separation along a circular array}
\label{app:separation}
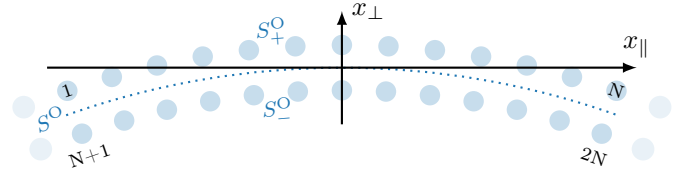
\begin{figure}[ht]

\begin{tikzpicture}[
	x=1cm,y=1cm,
	scale = 30]
	
	\def\nRecOuter{6}
	\def\NRecOuter{114}
	
	\def\phi{20}
	
	\def\RrecInner{0.3529cm}
	\def\RrecOuter{0.3729cm}
	
	\pgfmathsetlengthmacro{\RrecMid}{(\RrecInner+\RrecOuter)/2}
	
	\pgfmathsetlengthmacro{\RrecDist}{(\RrecOuter-\RrecInner)/2}
	
	\def\xn{0.025cm}
	\def\xt{0.13cm}
	
	\def\Rsource{0.005cm} 

\foreach \r in {\RrecInner,\RrecOuter}{
		\foreach \i in {-\nRecOuter,...,\nRecOuter}{
			\fill[\colrec,opacity=0.25] (90-360/\NRecOuter*\i:\r) circle [radius=\Rsource*0.9];
		};
	};

\pgfmathsetlengthmacro{\nRecFade}{\nRecOuter+1)}
	\foreach \r in {\RrecInner,\RrecOuter}{
		\foreach \i in {-\nRecFade,\nRecFade}{
			\fill[\colrec,opacity=0.1] (90-360/\NRecOuter*\i:\r) circle [radius=\Rsource];
		};
	};

	\pgfmathsetlengthmacro{\phi}{360/\NRecOuter*\nRecOuter}
\node[rotate=\phi] at (90+\phi:\RrecOuter) {\scriptsize 1};
	\node[rotate=-\phi] at (90-\phi:\RrecOuter) {\scriptsize N};
	
	\pgfmathsetlengthmacro{\RrecInnerIndices}{\RrecInner*0.97}
	\node[rotate=\phi] at (90+\phi:\RrecInnerIndices) {\scriptsize N+1};
	\node[rotate=-\phi] at (90-\phi:\RrecInnerIndices) {\scriptsize 2N};

\pgfmathsetlengthmacro{\phi}{360/114*(\nRecOuter+0.2)}
	\pgfmathsetlengthmacro{\xarc}{sin(\phi)*\RrecMid}
	\pgfmathsetlengthmacro{\yarc}{cos(\phi)*\RrecMid}
	\draw[dotted,\colrec,thick] (-\xarc,\yarc) arc [start angle=90+\phi,end angle=90-\phi, radius=\RrecMid] ;

\draw[thick,-latex] (-\xt,\RrecMid) -- (\xt,\RrecMid) node[above,anchor=south] {$x_\parallel$};
	\draw[thick,-latex] (0,\RrecMid-\xn) -- (0,\RrecMid+\xn) node[above,anchor=west] {$x_\perp$};

	\def\phi{20.7}
	\node[shift = {(0pt,0pt)},\colrec,font=\footnotesize,rotate=\phi] at (90+\phi:\RrecMid+\RrecDist*0.1) {$S^\text{O}$};
	
	\def\phi{4.7}
	\node[\colrec,font=\footnotesize,rotate=\phi,anchor=south] at (90+\phi:\RrecMid +\RrecDist*0.7) {$S^\text{O}_+$};
	\node[\colrec,font=\footnotesize,rotate=\phi,anchor=north] at (90+\phi:\RrecMid -\RrecDist*0.8) {$S^\text{O}_-$};

\end{tikzpicture} 	\caption{Construction of tangential, local coordinate surface where the origin coincides central point of the segment $S^\mathrm{o} \subset S^\mathrm{O}$.}
	\label{fig:arcShift}
\end{figure}

For acoustic cloning the required \gfs{} are obtained experimentally from the method of multidimensional deconvolution. This method relies on separating the wavefield on the recording surface $\so$ into ingoing and outgoing components ${\{p,v\} = \{p,v\}^\mathrm{in} + \{p,v\}^\mathrm{out}}$. Note that for \cref{eq:mdc} we only need the ingoing part.
Since these data are is recorded along closed, circular receiver arrays, we use a wavefield separation method for arbitrary curved, closed surfaces, as in \citet{liClosedapertureUnboundedAcoustics2021}. This requires either pressure recordings at two spatially offset levels or both the pressure and normal particle velocity recorded on a single surface. We use the two auxiliary surfaces $\so_\pm$ to record finite-difference pressure values to estimate the pressure and normal particle velocity on $\so$. In contrast to \citet{liClosedapertureUnboundedAcoustics2021} we modify the tangential axis of the local coordinate surface to coincide with $\so$ instead of $\so_-$ since we want the separated wavefield on $\so$.
\begin{figure}
	\centering

\begin{tikzpicture}[
	anchor=center,
	start chain=going below,
	every join/.style={arrow},
	node distance=0.4cm
	]
	
	\node [startstop, on chain,join] {Wavefield Separation};
	
	\node [action, on chain,join] {Record data on $\so$};
	
	\node [data, on chain,join] (illuminate) {Construct local coordinate system $S^\mathrm{loc}$ for a subset of the data around $x^\mathrm{o} \in \so$};

	\node [data, on chain,join]  {Fourier transform $\mathbf{p}_x(t) \overset{\mathcal{F}}\mapsto \mathbf{p}_x(\omega)$ and invert $\mathbf{p}_x(\omega)=\mathbf{F}\mathbf{p}_k(\omega)$ for $\mathbf{p}_k$.};
	
	\node [data, on chain,join]  {Inverse transform $\mathbf{p}_{+k}(\omega)\overset{\mathcal{F}^{-1}}\mapsto \mathbf{p}_x^\mathrm{in}(t)$ to get the ingoing field on the construction surface.};
	
	\node [data, on chain,join] {Retain $\mathrm{p}_{x^\mathrm{o}}^\mathrm{in}(t)$ for central point where construction surface coincides with $\so$.};

	\node [decision, on chain,join] (decide) {$o=N^\text{rec}$};

	\node [startstop, on chain,join] (stop) {Done};
	
\path [line] (decide) -- node[right]{yes}(stop);
	
	\draw [arrow,-] (illuminate.east) -- ++(0.5,0)  |-  (decide.east)node[above,xshift=1em]{no};
\end{tikzpicture} 	\caption{Wavefield separation method in the frequency-wavenumber domain.} \label{fig:flowwavefield separation}
\end{figure}
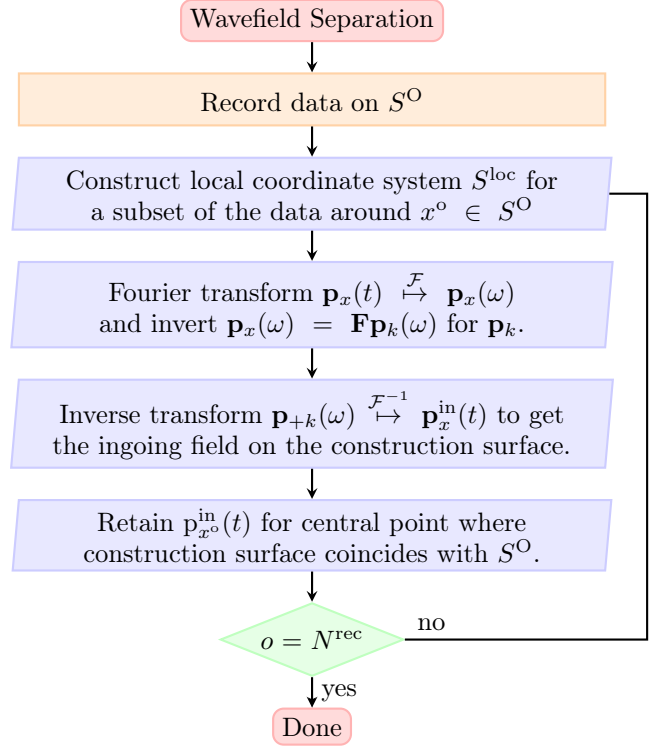
In \cref{fig:arcShift}, a local group of $2N$ neighboring receivers on a segment $S^\mathrm{o} \subset S^\mathrm{O}$ is shown. We chose $N$ odd such that there are a total of $(N-1)$ receivers to the left and right of the central receivers. For the above results $N=21$ was used. Around the midpoint $\mathbf{x}^\mathrm{o}\in \so$ of the chosen segment, a local coordinate system consisting of a normal and tangential axis to $S^\mathrm{o}$ is constructed. Furthermore, we index the microphones clockwise, from outer to inner auxiliary surface $S^\mathrm{o}_\pm=\{S^\mathrm{o}_+,S^\mathrm{o}_{-}\}$.
From this subset of pressure recordings $p(S^\mathrm{o}_\pm,t)$, we can separate the wavefield into its ingoing and outgoing components. Since at the origin the local coordinate system coincides with $\so$, we retain the separated wavefield for this location and proceed with processing every other point on $\so$ likewise.

First we transform $p(S^\mathrm{o}_\pm,t)$ to the frequency domain using the Fourier transform $p(\textbf{x},\omega)=\int p(\textbf{x},t) e^{-i\omega t} \,  dt$ to get $p(S^\mathrm{o}_\pm,\omega)$, with $i$ denoting the imaginary unit. Now we process each discretized frequency separately. The ingoing and outgoing components of the pressure in the wavenumber-frequency domain are then defined as $p(k_\parallel)=\{p^\mathrm{in},p^\mathrm{out}\}(k_\parallel)$. The tangential wavenumber $k_\parallel$ is discretized as $k^m_\parallel = (f/c)(m-N-1)/N$  for a regular spacing, with $\omega=2\pi f$, the acoustic wave speed $c$, and $m = 1,\ldots,2N$ such that we consider only propagating waves with $\mid k_\parallel \mid < f/c$. Defining the vector $p^n_\mathrm{x}=p(\mathbf{x}^n)$ and $p^m_\mathrm{k}=p(k^m_\parallel)$, we can relate them as
\begin{equation}
	p_\mathrm{x}^n=\mathrm{F}^{nm} p_\mathrm{k}^m,
	\label{eq:spaceFourier}
\end{equation}
with
\begin{equation}
		\mathrm{F}^{nm} = \exp [i2\pi(\, k^m_\parallel x_\parallel^n  + k_\perp^mx_\perp^n\,) ]
		\label{eq:spaceFourierOperator}
\end{equation}
and $k_\perp^m =s^m \sqrt{(f/c)^2-(k_\parallel^m)^2}$ the normal wavenumber as a function of the tangential wavenumber. The sign of the normal wavenumber is chosen positive for ${m<N}$ as ${s^m=\mathrm{sgn}(N-m)}$, ensuring the separation of the normal propagation direction \cite{ferberNoiseTransferVariabledepth2017}. On the left-hand side of \cref{eq:spaceFourier}, $\mathbf{p}_\mathrm{x}$ contains irregularly sampled pressure values in local 2D space, and on the right-hand side, $\mathbf{p}_\mathrm{k}$ denotes the pressure for a regularly sampled tangential wavenumber.

Relation \eqref{eq:spaceFourier} can be inverted in a least-squares sense using zero-order Tikhonov regularization wavefield $\mathbf{p}_k$ \cite{aster_parameter_2013}
\begin{equation}
	\mathbf{p}_\mathrm{k}\approx (\mathbf{F}^\mathrm{H} \mathbf{F} + \epsilon \mathbf{I} )^{-1}\mathbf{F}^\mathrm{H} \mathbf{p}_\mathrm{x},
	\label{eq:leastSquares}
\end{equation}
with $\mathbf{F}^\mathrm{H}$ the conjugate transpose of $\mathbf{F}$, the identity matrix $\mathbf{I}$, and a small-valued factor $\epsilon$ alleviating the problem of rank deficiency when inverting the Gram matrix $\mathbf{F}^\mathrm{H} \mathbf{F}$.

We are interested only in the field at the origin of the local coordinate system in space, $\mathbf{x}^\mathrm{o} = (0,0)$. Therefore, using \cref{eq:spaceFourier} and \cref{eq:spaceFourierOperator} we get $p^\mathrm{o,in}= \sum_{n=1}^N p_\mathrm{k}^n$. For the separated normal particle velocity, we additionally multiply with $k_\perp^m \,/(\rho f)$ in the wavenumber-frequency domain.

Since $p_\mathrm{o}^\mathrm{in}(\omega)$ is in the frequency-space domain, this is done for all discretized frequencies $\omega$. We then apply the inverse Fourier transform to get the time-domain signal $p^\mathrm{in}(\mathbf{x}_\mathrm{o},t)$ at the central point of this $\so$ segment (same for particle velocity). \cref{fig:flowwavefield separation} summarizes the wavefield separation algorithm. For more details, see \citet{liClosedapertureUnboundedAcoustics2021}.

\section{MDD and real-time holography}
\label{app:mdd}
The two steps of acoustic cloning, \cref{fig:flowclone}, are visualized here. First, we perform \gfs{} retrieval, \cref{fig:flowmdd}, followed by real-time broadband holography \cref{fig:flowholography}, to complete the cloning process.
\begin{figure}
	\centering

\begin{tikzpicture}[
	anchor=center,
	start chain=going below,
	every join/.style={arrow},
	node distance=0.4cm,
	]
	
	\node [startstop,on chain,join]	{Green's Functions Retrieval};

	\node [action, on chain,join] (illuminate) {Illuminate from $\phi^j$};
	
	\node [data, on chain,join] {Record wavefield $\{p,v\}$ on $\si$ and $\so$};
	
	\node [decision, on chain,join] (decide) {$j=N^\text{src}$};
	
	\node [data, on chain,join] (choosepoint) {For $\bm{x}^i$ on the inner surface $\si$};
	
	\node [data, on chain,join] (choosesrc) {For illumination source $\phi^j$};
	
	\node [data, on chain,join] {Separate wavefield on $\so$ $\{p,v\} \mapsto \{p,v\}_\text{out} + \{p,v\}_\text{in}$};
	
	\node [data, on chain,join] {Compose multidimensional convolution equation \eqref{eq:mdc}: $\{p,v\}_\text{in}(\so) \overset{G}\mapsto \{p,v\}(\bm{x}^i)$};
	
	\node [decision, on chain,join] (decidesrc) {$j=N^\text{src}$};
	
	\node [data, on chain,join] (mdd) {Solve multidimensional deconvolution for $G(\bm{x}^i\,|\,\xout)$};
	
	\node [decision, on chain,join] (decidepoint) {$i=N^\text{emt}$};
	
	\node [startstop, on chain,join] (stop) {Done};

\draw [arrow,-] (illuminate.west) -- ++(-0.25,0) |- (decide.west)node[above,xshift=-1em]{no};
	
	\path [line] (decide) -- node[right]{yes}(choosepoint);
	
	\draw [arrow,-] (choosesrc.west) -- ++(-0.25,0)  |- (decidesrc.west)node[above,xshift=-1em]{no};
	
	\path [line] (decidesrc) -- node[right]{yes}(mdd);
	
	\draw [arrow,-] (choosepoint.west) -- ++(-0.5,0)  |- (decidepoint.west)node[above,xshift=-1em]{no};
	
	\path [line] (decidepoint) -- node[right]{yes}(stop);
	
\end{tikzpicture} 	\caption{Green's function retrieval using MDD. Furthermore we have $G\equiv \{G^{p\mid q},G^{p\mid f},G^{v\mid q},G^{v\mid f}\}$, $N^\text{src}=52$ and $N^\text{emt}=18$.}
	\label{fig:flowmdd}
\end{figure}
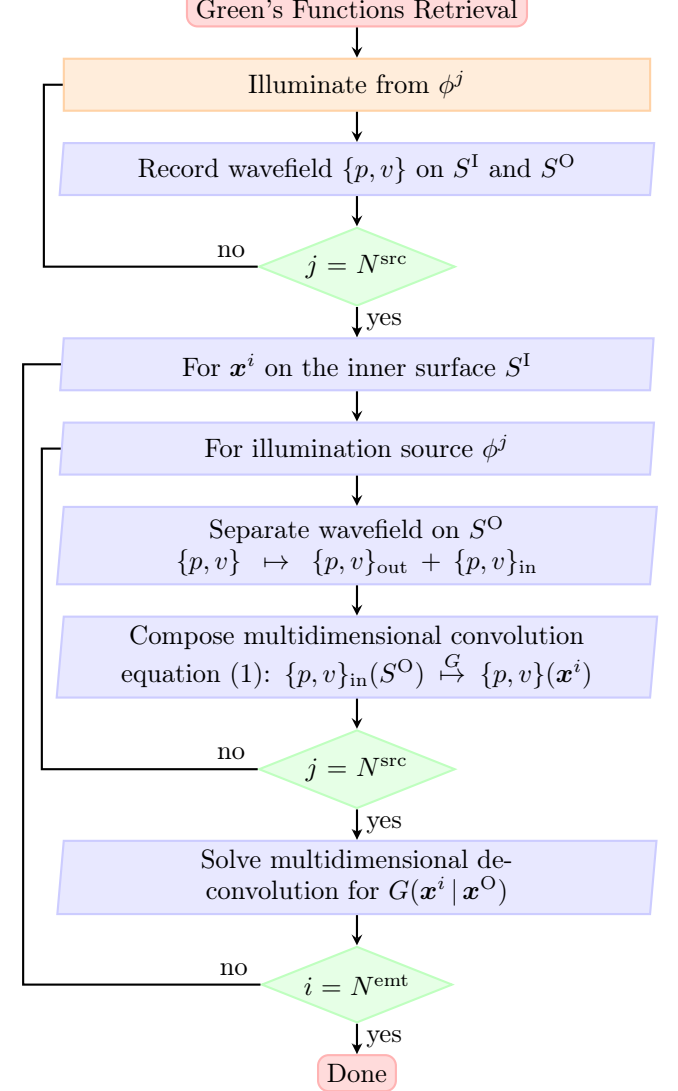
\begin{figure}
	\centering

\begin{tikzpicture}[
	anchor=center,
	start chain=going below,
	every join/.style={arrow},
	node distance=\baselineskip
	]
	
	\node [startstop,on chain,join] {Real-Time Holography};

	\node [data, on chain,join] (loadgf) {Load extrapolator kernels $G(\si \,|\, \so)$};
	
	\node [action, on chain,join] {Illuminate $V^\mathrm{O}$with arbitrary source field};
	
	\node [data, on chain,join] (rec) {Sample wavefield $\{p,v\}(\so,t_s)$};
	
	\node [data, on chain,join] {Extrapolate current field $\{p,v\}(\so,t_s) \overset{G}\mapsto \{p,v\}(\si,t\,|\, t_s)$};
	
	\node [data, on chain,join] {Add current extrapolation to buffer $\{p,v\}(\si,t\,|\, t_0,\dots, t_{s-1}) \pluseq \{p,v\}(\si,t\,|\, t_{s})$};
	
	\node [action, on chain,join] {Actuate monopole and dipole sources with current buffer values as source strengths ${\{q,f\}(\si,t_s) = \{v,p\}(\si,t_s)}$};
	
	\node [decision, on chain,join] (decidet) {$s=N^\text{T}$};
	
	\node [startstop, on chain,join] (stop) {Done};

	\draw [arrow,-] (rec.east) -- ++(0.5,0) |- (decidet.east)node[above,xshift=1cm]{$s\pluseq1$};
	
	\path [line] (decidet) -- node[right]{yes}(stop);
\end{tikzpicture} 	\caption{Real-time broadband holography. With $G\equiv \{G^{p\mid q},G^{p\mid f},G^{v\mid q},G^{v\mid f}\}$, $N^\text{src}=52$ and ${a\pluseq b}$ the addition assignment ${a= a+ b}$.}
	\label{fig:flowholography}
\end{figure}
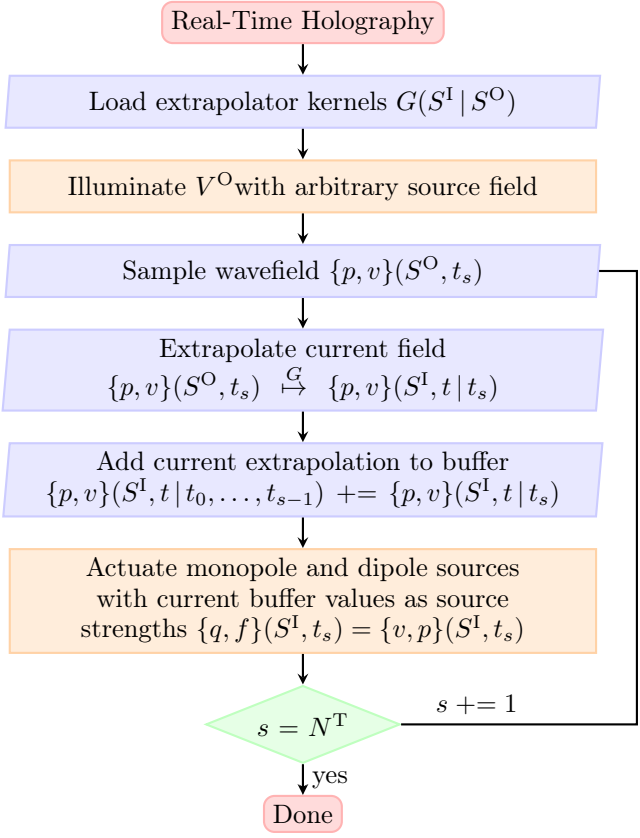

\section{Experimental setup}
\label{app:setup}
The Ricker wavelets we use are defined by $r(t)=\left(1-2(\pi f_c t)^2\right) \exp \left(- (\pi f_c t)^2\right)$ with $f_c=3$~kHz the center frequency. Using $c=343$~m/s the central wavelength is $\lambda_c=11$~cm.
 
The discretization of the integration surfaces reduces them to sums, with the question of accuracy arising. Considering the angular spacing of the inner ring, $d^\mathrm{I} = 2\pi r^\mathrm{I}/N^\mathrm{emt} = 3$~cm (the outer ring has an even denser discretization) is still considerably smaller than the central wavelength used and we therefore regard the discretization as sufficiently accurate.
 
Around the central frequency, the loudspeaker, with a diameter of $d_\mathrm{speaker} = 1.5$~cm, can be considered an acoustic monopole source. The monopole transducer then corresponds to a point source of volume injection rate $q$ \cite{hoopHandbookRadiationScattering1995}. We can therefore construct a dipole source by considering a pair of closely spaced monopole sources centered at $\si$ and aligned along the normal to the surface. Using an opposite sign ($180^\circ$ phase shift) on one component, we obtain an effective dipole centered on $\si$ in the low-frequency limit \cite{beranekAcousticsSoundFields2012}.

The auxiliary surfaces $\si_\pm$ are therefore equipped with loudspeakers to approximate effective monopole $q$ and normal dipole $f$ sources centered on $\si$. The monopole is obtained with an equal phase on both speakers ${q^i\approx \{p_+^i,p_-^i\}}$ while the dipole is achieved with opposite phase on the speakers $f^i\approx \{-p_+^i,p_-^i\}$. Note that this extends the spatial support of the monopole source due to the distance of the two auxiliary surfaces $d=\Delta r$. Nonetheless, we still work within the long-wavelength approximation disregarding the approximation errors.

\end{document}